# Monocrystalline Si/β-$Ga_2O_3$ p-n heterojunction diodes fabricated via grafting

Jiarui Gong[1,a)], Donghyeok Kim[1,a)], Hokyung Jang[1,a)], Fikadu Alema[2,a)], Qingxiao Wang[3,a)], Tien Khee Ng[3], Shuoyang Qiu[1], Jie Zhou[1], Xin Su[1], Qinchen Lin[1], Ranveer Singh[1], Haris Abbasi[1], Kelson Chabak[4], Gregg Jessen[4], Clincy Cheung[5],Vincent Gambin[5], Shubhra S. Pasayat[1], Andrei Osinsky[2, b)], Boon, S. Ooi[3, b)], Chirag Gupta[1,b)], Zhenqiang Ma[1,b)]

[1]*Department of Electrical and Computer Engineering, University of Wisconsin-Madison, Madison, Wisconsin, 53706, USA*

[2]*Agnitron Technology Incorporated, Chanhassen, MN 55317, USA*

[3]*Department of Electrical and Computer Engineering, King Abdullah University of Science and Technology, Thuwal 23955-6900, Saudi Arabia*

[4]*Air Force Research Laboratory, Dayton, OH 45433, USA*

[5]*Northrop Grumman, Redondo Beach, CA 90278, USA*

[a)] These authors contributed equally to this work.

[b)] Author to whom correspondence should be addressed. Electronic mail: mazq@engr.wisc.edu, cgupta9@wisc.edu, andrei.osinsky@agnitron.com, boon.ooi@kaust.edu.sa

## Abstract

The β-$Ga_2O_3$ has exceptional electronic properties with vast potential in power and RF electronics. Despite the excellent demonstrations of high-performance unipolar devices, the lack of p-type doping in β-$Ga_2O_3$ has hindered the development of $Ga_2O_3$-based bipolar devices. The approach of p-n diodes formed by polycrystalline p-type oxides with n-type β-$Ga_2O_3$ can face severe challenges in further advancing the β-$Ga_2O_3$ bipolar devices due to their unfavorable band alignment and the poor p-type oxide crystal quality. In this work, we applied the semiconductor grafting approach to fabricate monocrystalline Si/β-$Ga_2O_3$ p-n diodes for the first time. With enhanced concentration of oxygen atoms at the interface of Si/β-$Ga_2O_3$, double side surface passivation was achieved for both Si and β-$Ga_2O_3$ with an interface $D_{it}$ value of 1-3 × $10^{12}$ /$cm^2$·eV. A Si/β-$Ga_2O_3$ p-n diode array with high fabrication yield was demonstrated along with a diode rectification of 1.3 × $10^7$ at ±2 V, a diode ideality factor of 1.13 and avalanche reverse breakdown characteristics. The diodes C-V shows frequency dispersion-free characteristics from 10 kHz to 2 MHz. Our work has set the foundation toward future development of β-$Ga_2O_3$-based transistors.

## 1. Introduction

Beta-phase gallium oxide ($\beta$-$Ga_2O_3$), an ultrawide-bandgap semiconductor, has attracted substantial attention in recent years due to its exceptional electronic properties and its vast potential in power electronics, solar-blind optoelectronics, and even potentially in radio-frequency electronics [1, 2]. With an ultrawide bandgap of 4.9 eV [3], a high breakdown electrical field of 10.3 MV/cm [2], a high electron saturation velocity ($1.1 \times 10^7$ cm/s) [2], excellent thermal stability, and mature growth techniques [1], $\beta$-$Ga_2O_3$ is expected to generate impact in high-voltage, high-power, and high-temperature application. Various epitaxy techniques, including metalorganic chemical vapor deposition (MOCVD) [4-6], molecular-beam epitaxy [7, 8], and halide vapor phase epitaxy [9, 10], have been successfully demonstrated to produce high-quality epitaxial layers for device development. Furthermore, the easy availability of the native bulk substrates with a large diameter promises a lower cost through mass production in the industry in the future.

Despite these attractive properties of $\beta$-$Ga_2O_3$ and the excellent unipolar devices demonstrated, e.g., Schottky diodes [11-13] and field-effect transistors [14-17], there are some critical challenges to be addressed. For instance, the long-standing issue of lack of p-type doping in $\beta$-$Ga_2O_3$ has persisted. The p-type doping inefficiency stems from high ionization energy of acceptors [18, 19] when using the common dopants in $Ga_2O_3$. In the meantime, low hole mobility of merely around 41.4 $cm^2/V \cdot s$ in $\beta$-$Ga_2O_3$ [18] further hinders its potential application in high-speed electronics. As a result, the design and fabrication of high-performance bipolar $Ga_2O_3$ devices, such as p-n diodes, which are the building blocks of more complicated bipolar devices like bipolar transistors, are of vital importance in advancing the $\beta$-$Ga_2O_3$ related device technologies, however, the progress along this direction is still in the research and development stage.

There were some attempts in developing β-$Ga_2O_3$-based p-n diodes. Depositing p-type oxides on n-typ β-$Ga_2O_3$ showed p-n diodes with good rectification ratios [20-23]. However, the deposited p-type oxides are polycrystalline and the low crystal quality of these oxides limit the p-n diodes performance and would be infeasible for next-step triodes (e.g., transistors) device applications. To develop high-performance β-$Ga_2O_3$ based p-n diodes and other more complicated triode devices, monocrystalline p-type semiconductors like Si, GaAs, InP etc. are highly preferred.

In general, direct materials/wafer bonding/fusion of two monocrystalline semiconductors often leads to an interface layer having a high density of traps [24-43]. These traps serving as recombination and generation centers cause non-ideal rectification current-voltage (I-V) behavior in p-n junctions, including high ideality factors, high reverse-bias leakage current, difficulty to achieve avalanche breakdown and impossible to reach desired depletion width. As for β-$Ga_2O_3$, the wafer bonding of p-type diamond with n-type β-$Ga_2O_3$ showed limited device performance due to both high density of interfacial defects and unfavorable band alignment [44]. To overcome the issues associated with direct bonding/fusion, the semiconductor grafting approach [45] was introduced. The key feature of the grafting approach is that an ultrathin interfacial layer (*e.g.*, ultrathin oxide etc.) is inserted between two grafted semiconductors with arbitrary lattice structures (and lattice constants), which forms chemical bonding with both semiconductors and physically separate them. The major physical functions of the interfacial layer are double side passivation and effective quantum tunneling. The double-side passivation function passivates the dangling bonds of the semiconductors surfaces thereby reducing the surface density of states, and with the minimum recombination from the passivation, both electrons and holes can transport across the interface with negligible loss via quantum tunneling. Consequently, the resultant grafted junctions can function like lattice-matched ones formed by epitaxy.

In the following paragraphs, the design of grafted Si/β-$Ga_2O_3$ p-n junctions is described. First, the double-side passivation requirement in grafting is considered. As mentioned earlier, generally, an ultrathin oxide is inserted between two monocrystalline semiconductors to realize double-side passivation. In the case of β-$Ga_2O_3$, the interface ultrathin oxide can be $Al_2O_3$ or other types of oxides. However, the evidence of having almost no Fermi level pinning in previously demonstrated Schottky diodes [46, 47] strongly signifies that the surface of β-$Ga_2O_3$ is self-passivated. As a result, inserting a foreign ultrathin oxide like the $Al_2O_3$ to passivate the surface of β-$Ga_2O_3$ as practiced earlier [45] can be omitted in this case. By eliminating the extra foreign oxide, a thinner interface can be achieved to facilitate charge carrier transport across it. Moreover, according to the known band alignment between Si and $Ga_2O_3$, $Ga_2O_3$ can also passivate Si surface [2]. Therefore, the surface $GaO_x$ layer itself on top of the β-$Ga_2O_3$ is expected, to some extent, to fulfill the double-side passivation requirement, as needed for successful grafting between Si and β-$Ga_2O_3$. To enhance the double-side passivation of the $GaO_x$ layer, it is natural to further enrich the oxygen atoms of the surface of the β-$Ga_2O_3$, considering that it is the oxygen atoms that render the passivation functions.

Second, the design of interface for effective quantum tunneling is considered. When forming Si/β-$Ga_2O_3$ heterointerface the surface oxygen atoms of the β-$Ga_2O_3$ layer will react with the surface Si atoms to passivate the Si surface dangling bonds via a thermal process, which may lead to the possible formation of a $SiGaO_x$ layer at the interface. To make the interface $SiGaO_x$ layer meet the requirement of effective quantum tunneling for charge carriers transporting across the interface, the interfacial $SiGaO_x$ layer should be maintained as thin as possible. This requirement imposes a low thermal budget to be applied to the heterojunction grafting process in order to avoid an excessive atomic interdiffusion between the two semiconductors, which otherwise can lead to a thick interface layer. On the other hand, the planned thermal budget must be sufficient such that a strong chemical bonding can form between the Si and the β-

$Ga_2O_3$, enabling mechanical robustness of the finally grafted heterostructures. It is noted that the above Si/β-$Ga_2O_3$ interface design considerations to satisfy the requirements of both sufficient double-side passivation and effective quantum tunneling differ much from that of materials bonding with surface activation (*i.e.*, surface activated bonding (SAB)) [48].

Our experiments were performed accordingly. First, a β-$Ga_2O_3$ epi substrate was grown. Then the surface of the β-$Ga_2O_3$ substrate was oxygen-enriched. We used X-ray photoelectron spectroscopy (XPS) analysis to verify the surface oxygen enrichment condition and study the surface atomic bonding conditions of the β-$Ga_2O_3$. A monocrystalline Si membrane was fabricated and transferred to the β-$Ga_2O_3$ substrate to form Si/β-$Ga_2O_3$ heterojunctions, followed by the Si/β-$Ga_2O_3$ p-n diode array fabrication. Scanning tunneling electron microscopy (STEM) analysis was carried out to examine the grafted Si/β-$Ga_2O_3$ interface. The p-n diodes were also characterized via I-V and capacitance-voltage (C-V) analysis to understand the interface density of states ($D_{it}$) of the grafted heterostructures.

## 2. Results and Discussion

Figure 1 shows the detailed information of the β-$Ga_2O_3$ epi substrate. Figure 1(a) shows an optical image of a piece of diced β-$Ga_2O_3$ epi substrate (15 × 10 $mm^2$) with its crystal angle illustrated. The full-width of the half maximum (FWHM) of the XRD 2 theta-omega scan of β-$Ga_2O_3$ (020) peak is 144" (Figure 1(b)). The root-mean-square surface roughness ($R_q$) measured by AFM in a scanning area of 2 × 2 $\mu m^2$ is 0.76 nm (Figure 1(c)). The smoothness of the surface of the β-$Ga_2O_3$ facilitates the grafting process with Si.

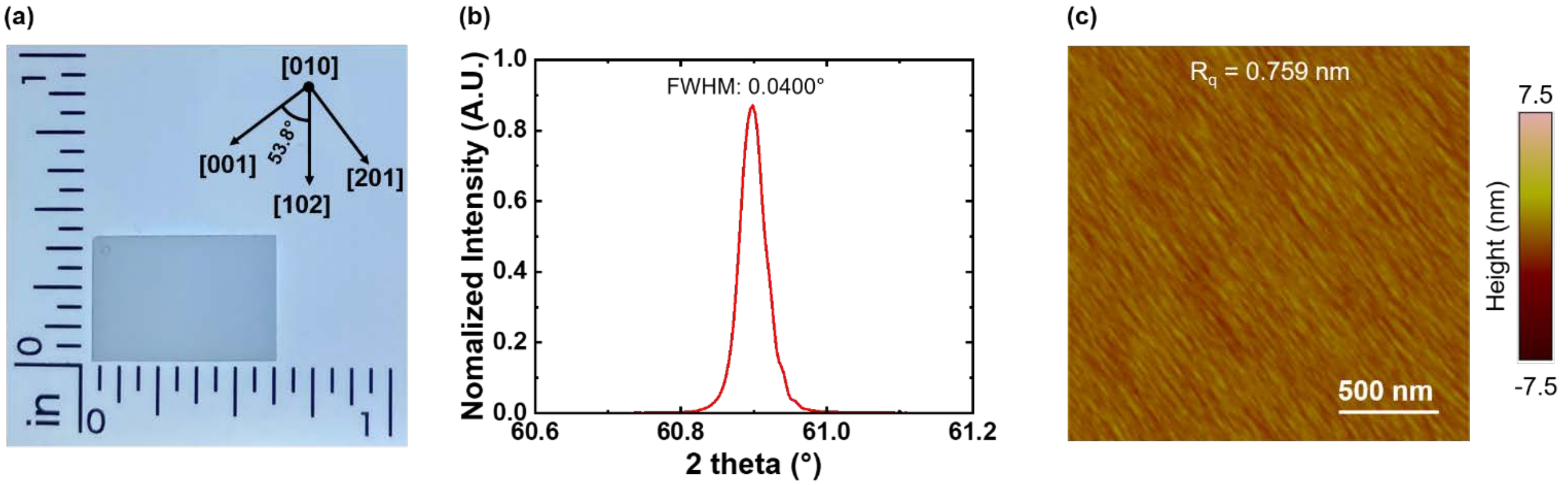

**Figure 1.** (a) An optical image of the β-$Ga_2O_3$ epi sample. (b) The 2 theta-omega XRD spectrum of the (002) peak from the as-grown β-$Ga_2O_3$ epi sample. (c). An AFM image taken on the epi surface (2 x 2 μ$m^2$) of the β-$Ga_2O_3$ sample. The $R_q$ value is 0.76 nm.

Figure 2(a) shows an optical image of a patterned p+ SOI sample with hole arrays dry-etched through the top Si layer. The center square area to be released as membrane is 2.6 × 2.6$mm^2$. The hole size is 9 × 9 μ$m^2$, as shown in the zoomed-in optical microscopic image. The blue area is BOX, which is exposed by RIE etching. The outer frame Si area (with of 3.8 mm × 3.8 mm) without holes serve as a square anchor, which is not released by HF, with three Si ribbons (300 μm × 300 μm) connecting each of the four sides of the square anchor to the center Si membrane. Figure 2(b) shows an image of the p+ Si membrane released inside the HF solution. The undercut process by HF solution can remove the BOX layer underneath the Si membrane area and the ribbons. Figure 2(c) shows the p+ Si membrane sitting on its original Si handling substrate after the HF release process was completed. Figure 2(d) shows an optical image of the picked Si membrane sitting on a PDMS stamp. The view of the surface is from the backside (which was in contact with the BOX layer before the undercut process) of the membrane. Figure 2(e) shows an AFM image of the backside of the Si NM sitting on the PDMS stamp with an $R_q$ of 0.15 nm. Figure 2(f) shows the XRD 2 theta-omega spectrum of a Si membrane transferred to a β-$Ga_2O_3$ substrate, from which the monocrystalline features of both Si and β-$Ga_2O_3$ can be clearly observed. The Raman spectra of both the Si/β-$Ga_2O_3$ stack and the standalone β-$Ga_2O_3$ substrate are shown in Figure 2(g). The Si peak was observed at 517.2 $cm^{-1}$.

Bulk Si [49] and β-$Ga_2O_3$ [50] have a large difference of thermal expansion coefficients. As the Si thickness is reduced, it can accommodate more mechanical strain [51] than a bulk substrate. In addition, Si membrane is very flexible and much more compliant than its rigid counterpart and can be fabricated into any sizes, ranging from tens of microns to inches, for transfer printing. As a result, monocrystalline Si membrane is considered more advantageous than bulk Si when forming heterojunctions between Si and β-$Ga_2O_3$.

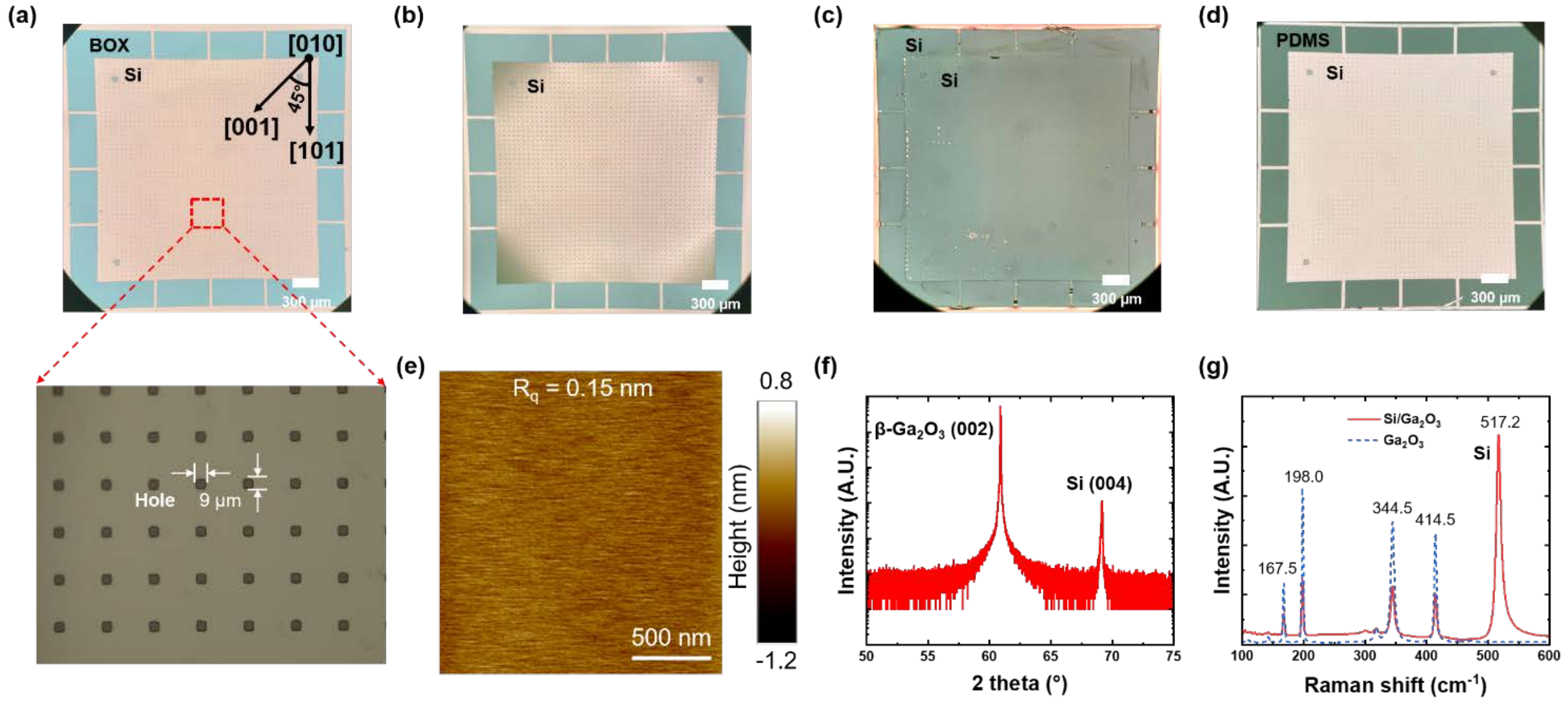

**Figure 2.** (a)-(d) Optical microscopic images of p+ Si membrane fabrication. (a) An image of a patterned p+ SOI substrate with hole arrays formed by RIE along with a magnified view to show the etched holes. (b) An image of the patterned SOI being released inside HF solution. The center Si membrane was floating in the HF solution while being anchored by the edge frame. (c) An image of the p+ Si membrane sitting on its handling Si substrate right after the release process was completed. The p+ Si membrane remains

flat. (d) An image of the p+ Si membrane attached to a PDMS stamp after being picked up. (e) The AFM image of the backside of the p+ Si membrane with a $R_q$ value of 0.15 nm. (f) XRD spectrum of a p+ Si membrane transferred to a β-$Ga_2O_3$ substrate. (g) Raman spectra of the Si/β-$Ga_2O_3$ stack (red solid) and the standalone β-$Ga_2O_3$ substrate (blue dash).

Figure 3(a) shows an optical microscopic image of the fabricated Si/β-$Ga_2O_3$ p-n diode array along with the schematic illustration of the vertical layer structures (Figure 3(b)). The fabricated diode array was cleaved into two halves along the (001) plane. The entire array consists of 336 diodes, all having the same anode metal size, and the fabrication yield is close to 100%. No visible Si membrane or anode metal peeling off was observed. A detailed view of one individual diode is shown in Figure 3(c).

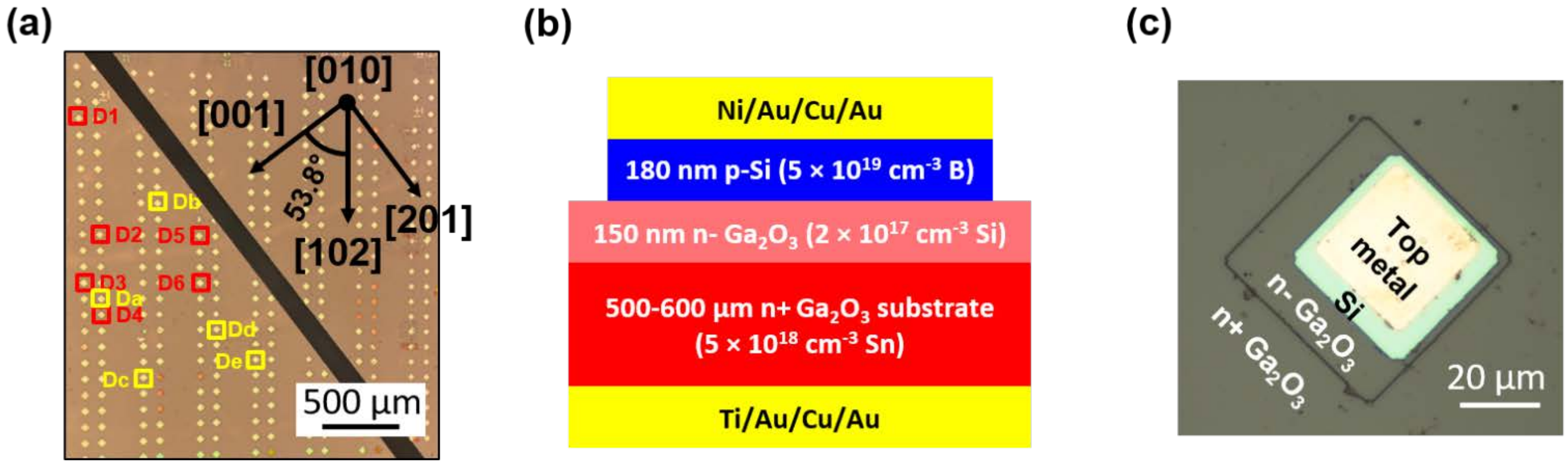

**Figure 3.** (a) An optical microscopic image of the fabricated Si/β-$Ga_2O_3$ p-n diode array consisting of 336 diodes. The labeled diodes were randomly selected for I-V and reverse breakdown characterizations. The crystal orientation of the β-$Ga_2O_3$ epi was illustrated. (b) The schematic illustration of the vertical layer structure and (c) a detailed view of an individual Si/β-$Ga_2O_3$ p-n diode.

Figure 4(a) depicts the measured I-V characteristics of randomly selected (see Figure 3(a) for the device labeling) Si/β-$Ga_2O_3$ p-n diodes in the range of -2 to 2 V. From an I-V measurement survey of the p-n diodes over an area of about 1 × 1 $mm^2$, the yield of the working device is estimated to be 86%. The ideality factors ($n$) of the Si/β-$Ga_2O_3$ p-n diodes shown in Figure 4(a) were calculated and listed in Table 1. According to the equation of $n$ as a function of interface defect density ($D_{it}$) in grafted heterojunction p-n diodes [45] and the interfacial layer thickness of 2.5 nm (to be described later), the $D_{it}$ values of the Si/β-$Ga_2O_3$ interface are estimated to be 1-3 × $10^{12}$ /$cm^2$·eV, considering that the interface dielectric constant may be any value in the range of 3.9 ($SiO_2$ [8]) to 10.43 (β-$Ga_2O_3$ [9]). The $D_{it}$ value is lower than that of the Si/β-$Ga_2O_3$ n-n diode fabricated by SAB [52]. As shown in Figure 4(a), the ultralow reverse current also indicates a very low density of generation centers at the grafted Si/β-$Ga_2O_3$ interface, making the grafted Si/β-$Ga_2O_3$ p-n diodes ideal for applications such as high-power devices and photo detectors, etc. Figure 4(b) further shows the measured breakdown characteristics of a few randomly selected devices (see Figure 3(a)). The breakdown ranges from 20-25 V (Table 2). This variation is because the avalanche phenomenon is a stochastic process.

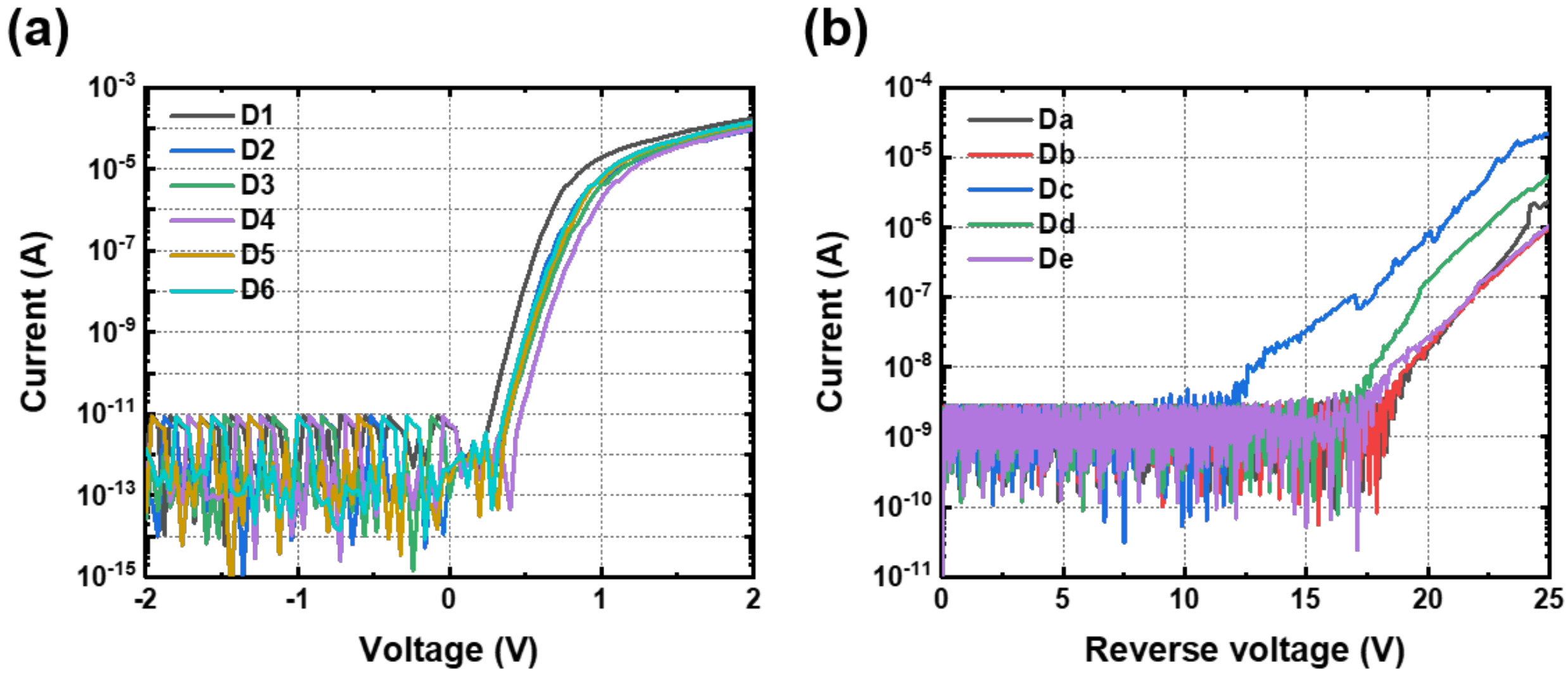

**Figure 4.** (a) I-V characteristics obtained from randomly selected Si/β-$Ga_2O_3$ p-n diodes in the range of ± 2 V. (b) Measured reverse I-V and breakdown characteristics of a few randomly selected Si/β-$Ga_2O_3$ p-n diodes.

**Table 1.** Summary of the ideality factor (*n*) of randomly selected Si/β-$Ga_2O_3$ p-n diodes.

| Device # | D1 | D2 | D3 | D4 | D5 | D6 |
|---|---|---|---|---|---|---|
| *n* | 1.13 | 1.17 | 1.36 | 1.26 | 1.24 | 1.22 |

**Table 2.** Summary of the breakdown voltage (BV) of randomly selected Si/β-$Ga_2O_3$ p-n diodes.

| Device # | Da | Db | Dc | Dd | De |
|---|---|---|---|---|---|
| BV (V) | 24.2 | 25 | 20.7 | 22.4 | 24.9 |

The measured C-V characteristics of the Si/β-$Ga_2O_3$ p-n didoes are shown in Figure 5. Little frequency dispersion was observed in the C-V curves in a frequency range of 10 kHz to 2 MHz (Figure 5(a) and 5(b)). The p-n junction capacitance values only vary with the reverse bias values that were applied to the diodes. The non-dispersion of the C-V curves indicates that a Si/β-$Ga_2O_3$ interface of high quality and low $D_{it}$ was attained. To estimate the upper values of the $D_{it}$ based on the C-V curves, Silvaco Atlas simulations of the Si/β-$Ga_2O_3$ p-n diode structure (Figure 3(b)) with different $D_{it}$ values were performed. The simulations results are shown in Figure 5(c). As can be seen, without significant change of capacitance within the given frequency range, the upper value of the $D_{it}$ is $6 \times 10^{12}$ /cm$^2$·eV, in agreement with the I-V measurement results.

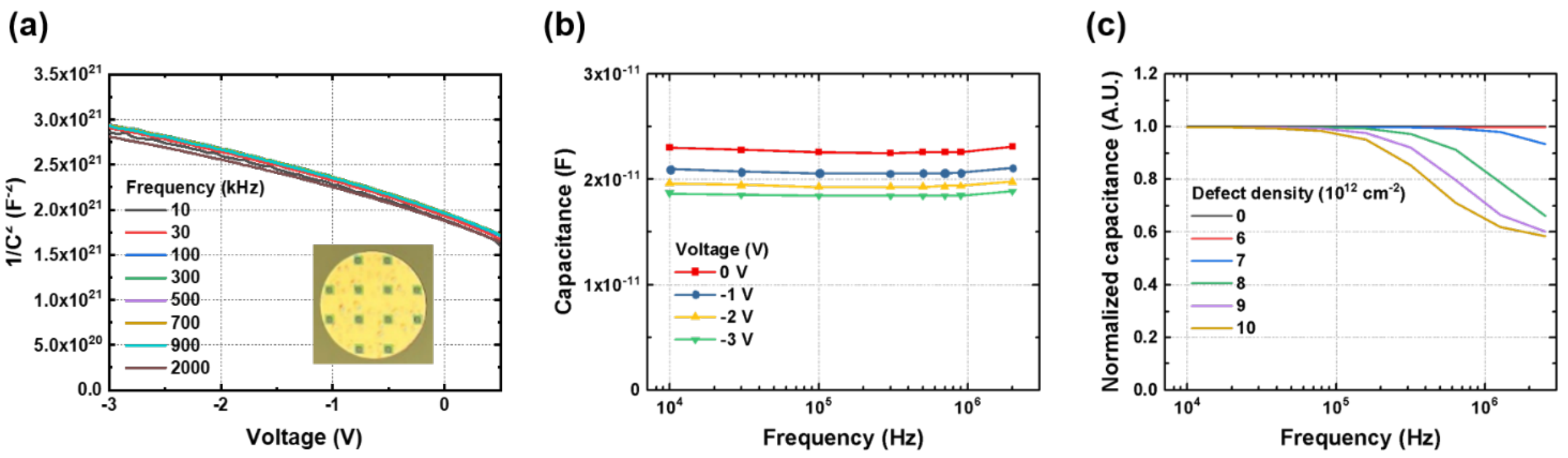

**Figure 5.** (a) Measured $1/C^2$ of the Si/β-$Ga_2O_3$ p-n diode as a function of bias voltage at multiple frequencies from 10 kHz to 2 MHz. The inset is the optical image of the C-V diode, which has a larger anode area. (b) Plot of the junction capacitance of the identical Si/β-$Ga_2O_3$ p-n diode in (a) as function of measurement frequency under different bias values from 0 to -3 V, showing that the diode junction capacitance only varies with the bias. (c) The simulated C-V plots of the Si/β-$Ga_2O_3$ p-n diode under 0 V with different interface $D_{it}$ values.

Considering the low $D_{it}$ values attained at the Si/β-$Ga_2O_3$ p-n junction, the Fermi level pinning due to charge traps at the grafted Si/β-$Ga_2O_3$ interface is expected to be negligible. Therefore, the band alignment between the Si and the β-$Ga_2O_3$ is would follow the electron affinity rule. Since the electron affinity of β-$Ga_2O_3$ is 4.00 eV [53], the Si/β-$Ga_2O_3$ p-n junction alignment can be constructed as shown in Figure 6. The Fermi energy level of the Si and that of the β-$Ga_2O_3$ were calculated according to the parameters shown in Fig. 3(b). Such an ideal band alignment can only be possible if the interface $D_{it}$ is sufficiently low [45]. It is noted that such a type-I band alignment between Si and β-$Ga_2O_3$ would be much more beneficial for future development of triode bipolar junction transistors than other β-$Ga_2O_3$-based pn junctions that employ the p-type oxides [20, 21, 23].

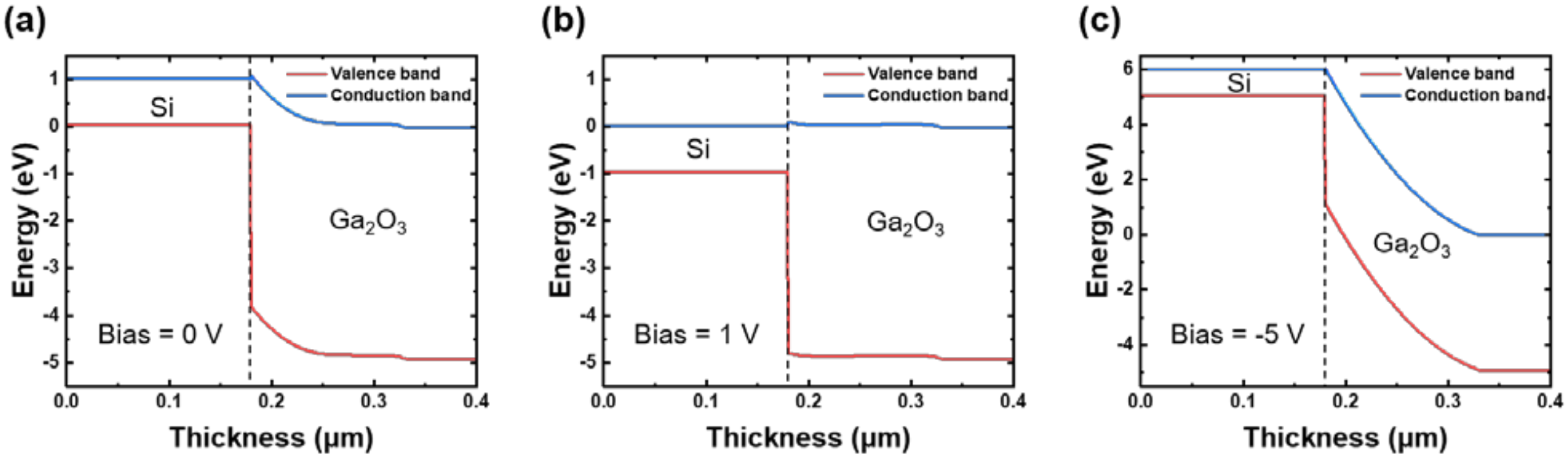

**Figure 6.** Constructed Si/β-$Ga_2O_3$ p-n junction band diagram under (a) equilibrium (0 V), (b) forward bias of 1.0 V, and (c) reverse bias of -5 V.

Figure 7(a) shows the drawing about the relative orientation of the monocrystalline Si and β-$Ga_2O_3$ of the characterized Si/β-$Ga_2O_3$ p-n diode. Figure 7(b) shows the overview (or low magnification) HAADF image of the grafted Si/β-$Ga_2O_3$ interface. The high-resolution (HR)-HAADF images viewed along two different lateral directions (Figure 7(c) and 7(d)) clearly show the β-$Ga_2O_3$ lattice structure along the [201] axis and the Si lattice structure along the [101] axis, respectively. The interfacial amorphous layer is estimated to be only ~2.0-2.5 nm thick. The thin interface layer ($SiGaO_x$) is expected to facilitate charge carrier transport through the interface. The small mis-orientation angle between the Si [101] and the β-$Ga_2O_3$ [201] axes was caused by “misalignment” during processes such as SOI wafer dicing and possibly during the Si membrane transfer. The mis-orientation however does not have any adverse effects on the device performance since the interface layer between Si and β-$Ga_2O_3$ is amorphous. In fact, the alignment applied when transferring Si membrane to β-$Ga_2O_3$ was merely for the ease of photolithography steps. No accurate alignment was needed or enforced during the Si membrane transfer.

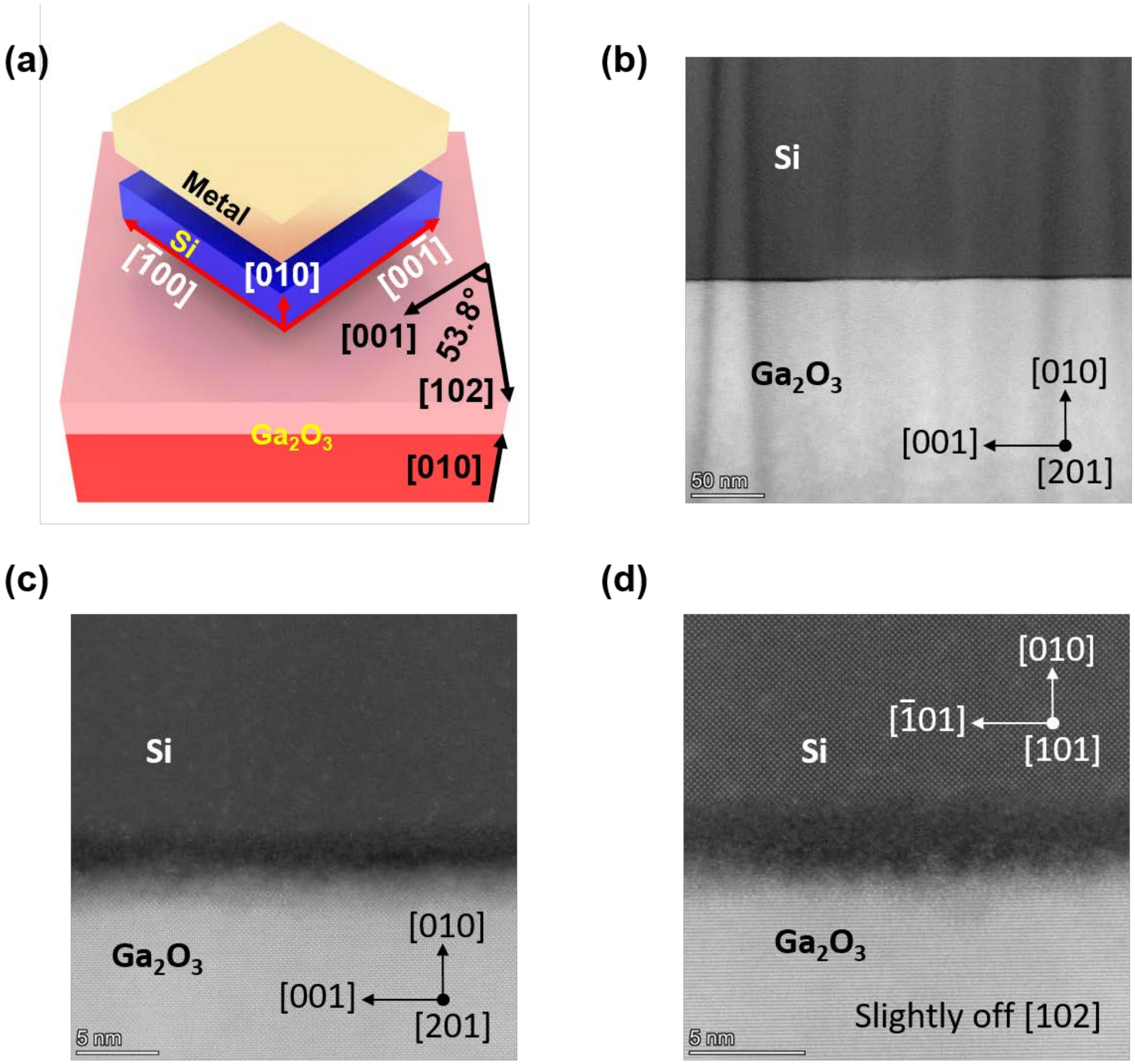

**Figure 7.** STEM imaging of the Si/β-$Ga_2O_3$ grafted interface. (a) Illustration of crystalline orientations of the characterized functional Si/β-$Ga_2O_3$ p-n diode. (b) A HAADF image of the Si/β-$Ga_2O_3$ interface. (c) and (d) HR-HAADF images showing the cross-sectional view of the Si/β-$Ga_2O_3$ interface with different viewing angles to show the lattice structure of the β-$Ga_2O_3$ and the Si, respectively.

Figure 8(a) shows the EELS scanning area of the Si/β-$Ga_2O_3$ interface for identifying the elements distribution, viewed along Si [101] axis and slightly off β-$Ga_2O_3$ [102] axis. Figures 8(b)-(e) show the false-color element distributions of Si, Ga, O, and C atoms across the interface and Figure 8(f) shows the overlay of all above distributions. By comparing the Si (Figure 8(b)) and the Ga (Figure 8(c)) distributions, Si and Ga are not intermixed. The clear physical separation of the Si atoms and the Ga atoms is critical in reducing the value of $D_{it}$ at the interface. Figure 8(d) indicates the presence of oxygen atoms at the interface. It is speculated that the oxygen atoms at the surface of the β-$Ga_2O_3$ reacted with the surface Si atoms during the thermal annealing process. This Si-O reaction passivates the surface dangling bonds of the monocrystalline Si. Combined with the self-passivation effect of the β-$Ga_2O_3$ surface, a substantially low $D_{it}$ value at the Si/β-$Ga_2O_3$ interface was obtained, as shown in I-V and C-V measurements. From the analysis, sufficient oxygen atom presence at the surface of β-$Ga_2O_3$ is essential to achieve adequate surface passivation of the Si surface, which perhaps also be advantageous for further improved surface passivation of β-$Ga_2O_3$. Figure 8(e) shows the presence of some C atoms at the interface, which was unexpected. However, by XPS analysis (to be detailed below), it was confirmed that C atoms were highly possibly from isopropyl alcohol (IPA) residue during the Si NM release process (Figure 10(b)). Furthermore, the effects C atoms on the Si/β-$Ga_2O_3$ p-n diode characteristics are unknown.

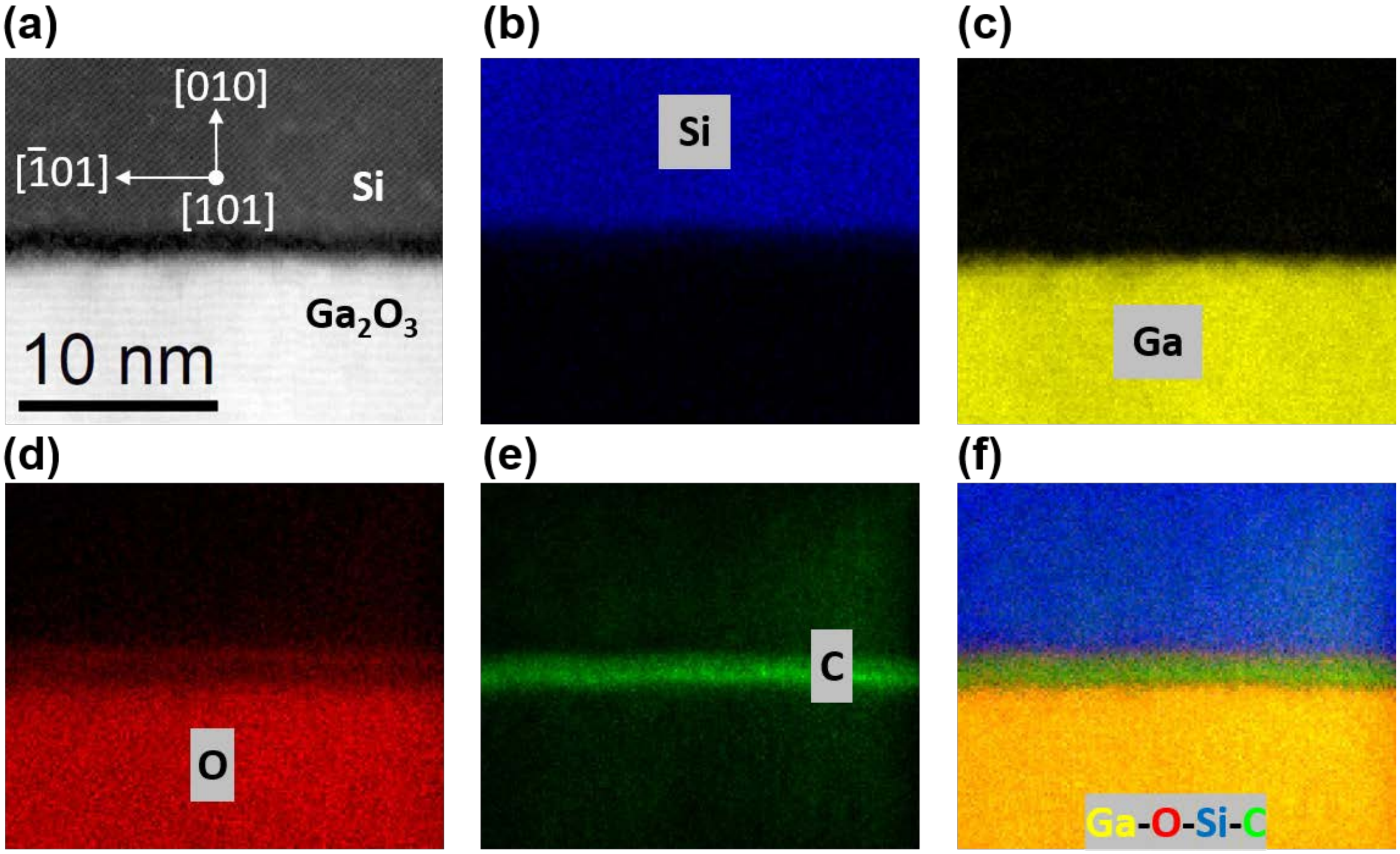

**Figure 8.** EELS elemental mapping at the Si/β-$Ga_2O_3$ interface. (a) The scanning region of the EELS mapping. (b)-(e) The fake color images showing the distributions of the element Si, Ga, O and C, respectively. (f) Overlaid distribution of Ga-O-Si-C elements at the interface.

As discussed above, the oxygen atoms at the interface of Si/β-$Ga_2O_3$ play the central role in achieving double side passivation of the two materials. To better understand the role of oxygen atoms on the surface of the β-$Ga_2O_3$ epi sample during its surface preparations, and particularly to understand how ion bombarding (*e.g.*, used in surface preparation in SAB) can affect β-$Ga_2O_3$ surface oxygen atoms, the XPS analysis results on the O 1*s*, Ga 3*d*, and C 1*s* spectra were illustrated in Figure 9. The measurements were taken from the β-$Ga_2O_3$ epi sample under three different surface conditions. The XPS measurement was first taken on the "grafting ready" sample, followed by two iterations of a 30-s Ar ion in-situ etching and an XPS measurement on the same sample. During the XPS spectra collection, auto-height process of the sample stage inside the XPS chamber was imitated at the beginning of the experiment. Hence, the absolute intensity of the spectra collected under the three surface conditions can be compared quantitatively. As clearly shown in Figure 9(a), the surface oxygen atom concentration has been reduced substantially (~5% reduction in peak area) by the first 30-s Ar ion etching. Since the Ar etching is typically used in a SAB process, the degradation of surface passivation due to the removal of the surface oxygen atoms should be inevitable. In our work, the enrichment of surface O atoms due to the soaking process of the β-$Ga_2O_3$ sample in the Piranha solution becomes the key step (opposing the surface treatments in SAB) to achieve adequate surface passivation for Si and β-$Ga_2O_3$. On the other hand, the concentration of surface Ga atoms increased (~3-4% increase in peak area) after the first 30-s Ar etching (Figure 4(b)). Furthermore, no significant change in XPS spectra was observed between the 30-s Ar-etched sample and the 60-s Ar-etched sample, indicating that the observed difference in Ga and O concentration only exists in a nanometer-scale surface layer and is not caused by the potential difference in sputtering rate between Ga and O atoms but real. Both Figure 9(a) and 9(b) indicate that 30-sec Ar ion etching is long enough to modify the surface of the "grafting ready" sample surface. Additional Ar ion beam etching does not induce further changes in the stoichiometry of the surface of the β-$Ga_2O_3$. Figure 9(c) shows that there was no carbon residue on the surface of the β-$Ga_2O_3$ sample (to the instrumental resolution limit of the XPS), which confirmed that the C source was from the Si side (*i.e.*, IPA). The XPS spectra suggests that there exists an ultrathin oxygen-rich layer on the surface of the β-$Ga_2O_3$ substrate. The oxygen atoms in this layer reacted with the surface atoms of the Si layer forming the final interface layer (~2-2.5 nm) that serve as the quantum tunneling the double-side passivation interface between the two monocrystalline crystals: the Si and the β-$Ga_2O_3$. The role of the interface $SiGaO_x$ has served the identical roles in other grafted heterostructures where an ultrathin $Al_2O_3$ layer was applied using atomic layer deposition (ALD) [45, 54-61].

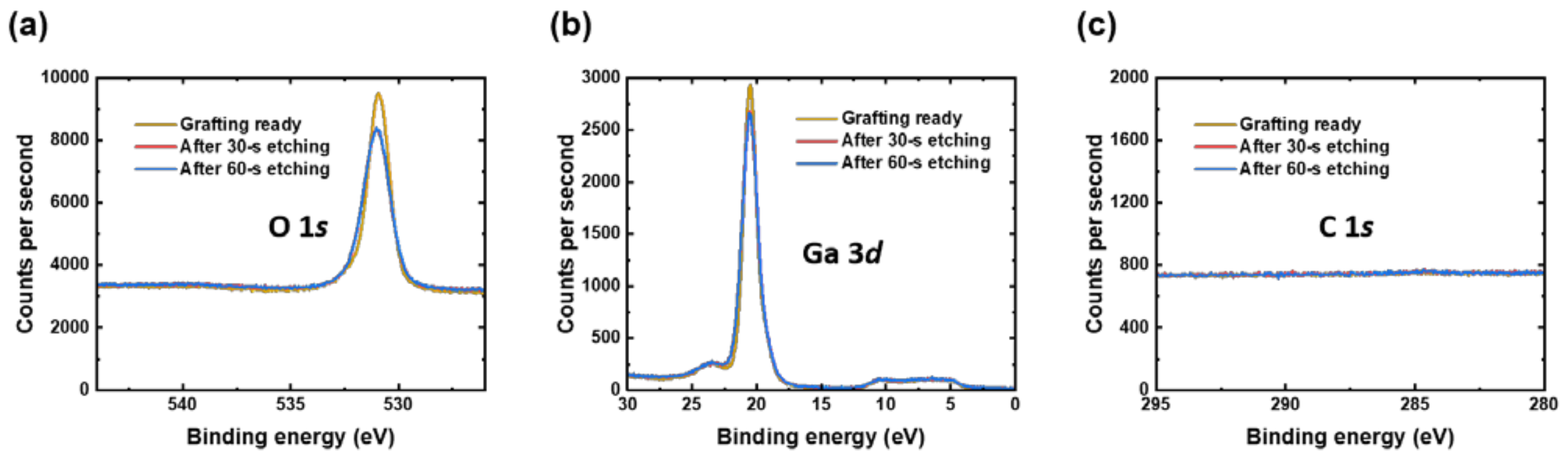

**Figure 9.** (a) Ga 3*d*, (b) O 1*s*, and (c) C 1*s* XPS spectra collected on the β-$Ga_2O_3$ epi sample under different surface conditions. The gold color spectra were collected from the "grafting ready" sample and the red color and the blue color spectra were collected after a 30-s and another 30-s ion-gun etching.

## 3. Conclusions

In this work, we applied the semiconductor grafting approach to fabricate Si/β-$Ga_2O_3$ p-n diodes for the first time. With enhanced concentration of oxygen atoms at the interface of Si/β-$Ga_2O_3$, sufficient double side surface passivation was achieved for both Si and β-$Ga_2O_3$ with a $D_{it}$ value of 1-3 × $10^{12}$ /$cm^2$·eV. By controlling the thermal budget during the chemical bonding formation process, ~2-2.5 nm interface amorphous layer was attained without inducing Ga and Si atoms inter-diffusion, verified by STEM imaging, to ensure effective quantum tunneling. A diode array with high fabrication yield was demonstrated along with a p-n diode rectification of 1.3 × $10^7$ at ±2 V, a diode ideality factor as low as 1.13 and avalanche reverse breakdown characteristics. The dispersion-free C-V versus frequency of the Si/β-$Ga_2O_3$ p-n junction confirmed the low $D_{it}$ results from the grafting approach. The near-ideal Si/β-$Ga_2O_3$ heterojunction formed in this work, the grafting approach being applied to combine single crystal Si as the p-type semiconductors with the n-type β-$Ga_2O_3$, and the membrane transfer printing method used in lieu of bulk Si (or SOI) all could be further expanded to the future development of β-$Ga_2O_3$-based transistors.

## 4. Experiment Section

*Epitaxy growth of β-$Ga_2O_3$*

Epitaxy growth of β-$Ga_2O_3$ and cleaning prior diode fabrication. The β-$Ga_2O_3$ epitaxial structure consists of ~150 nm thick lightly Si doped (~$2 \times 10^{17}$ $cm^{-3}$) homoepitaxial layer grown using Agnitron Technology's Agilis 500 MOCVD reactor [4-6] on a Sn-doped (~$5 \times 10^{18}$ $cm^{-3}$) (010) β-$Ga_2O_3$ substrate (Novel Crystal Technology) of ~550 µm. (Fig. 1(a) and 1(b)). Trimethylgallium (TMGa), pure oxygen (5N), and silane diluted in nitrogen ($SiH_4/N_2$) were used as precursors and argon (5N) as carrier gas during the growth. The Ar and $O_2$ gases were passed through point-of-use purifiers to reduce the impurity level to <1 ppm. Prior to loading the sample into the growth chamber, the substrate was cleaned using hydrofluoric acid (HF) for 30 minutes. The as-grown β-$Ga_2O_3$ epi sample was characterized by X-ray diffraction (XRD) and the surface roughness of the n-side surface of the β-$Ga_2O_3$ was characterized by atomic force microscope (AFM).

*Characterization of β-$Ga_2O_3$ substrate*

The as-grown β-$Ga_2O_3$ substrate was first dipped into acetone and sonicated for ten minutes, and the same process was repeated using isopropyl alcohol (IPA). After that, the substrate was put into Piranha solution (96% $H_2SO_4$ : $H_2O_2$ = 3:1) for ten minutes at room temperature. Then, the samples were thoroughly rinsed with deionized (DI) water for one minute. With this step, it is expected to enrich the oxygen atoms at the surface of the β-$Ga_2O_3$ substrate considering that oxygen atoms provide passivation function at the interface between Si and β-$Ga_2O_3$. The same cleaning procedures were iterated with RCA1 solution (29% $NH_4OH$ : 30% $H_2O_2$ : DI water =1:1:5) and RCA2 solution (37% HCl : DI water = 1:1) in sequence, the sample was dipped into a diluted HF solution (49% HF : DI water = 1:20) for one minute. This sample was named the "grafting-ready" sample.

The grafting-ready sample was performed using a Thermo Scientific K Alpha X-ray Photoelectron Spectrometer (XPS) with Al $K_{\alpha}$ X-ray source ($hv$ = 1486.6 eV). The sample was carried in a zip-bag filled with $N_2$ during the transportation from the lab to the XPS instrument to minimize the influence of air. It took about 10 min from moving the sample out of the HF solution to loading it into the XPS chamber. The instrument was calibrated by Au $4f_{7/2}$ peak at 84.0 eV. The following settings were applied to the spectrometer: 10 eV pass energy, 400 μm spot size, 5 s dwell time, and 0.05 eV step size. The auto-height of the sample stage was done only once at the start of the experiment. The Ga 3*d*, O 1*s*, and C 1*s* peaks were measured to collect the first set of data. The second set of data was obtained following a 30-s ion gun etching with 1 keV ion energy and a 2 mm raster size. And the third set of data was collected after another 30-s ion-gun etching.

*Fabrication of transferable p+ Si membranes*

To prepare Si membranes, a 6-inch Soitec$^{(R)}$ p-type (14-22 Ω·cm) silicon (100)-on-insulator (SOI) substrate was used to begin with. The 8-inch SOI substrate has a 205 nm Si device layer and a 400 nm buried oxide (BOX) layer. A thermal oxidation process using a furnace tube under 1050 °C was applied to the SOI substrate for 12 min to grow 36 nm thick screen oxide. Then the SOI substrate was boron implanted at room temperature with a dose of $3 \times 10^{15}$ cm$^{2}$ under 15 keV at an incident angle of 7°. After finishing the ion implantation, a thermal activation process at 950 °C for 40 min was performed using the same furnace tube. After finishing the activation process, the resultant Si device layer has a final thickness of 180-185 nm thick and a boron doping concentration ranging from $9.7 \times 10^{19}$ cm$^{-3}$ at its top surface to $9.5 \times 10^{19}$ cm$^{-3}$ at the bottom (next to the BOX layer), and the final top screen oxide become ~52 nm thick. All the above process conditions were guided with Silvaco Athena simulations, and after the dopant activation process the p+ Si device layer maintains its monocrystalline properties.

To fabricate Si membranes, the p+ SOI substrate was diced into 5-mm-by-5-mm squares. The size of dicing was to match the size of the β-$Ga_2O_3$ substrate used in this study. First, the top 52 nm thick screen oxide layer of the diced SOI squares was first removed by buffered oxide etchant solution. After that, a conventional photolithography process was performed to form a 9-μm-by-9-μm square-hole array pattern on the p+ Si device layer (Figure 10(a)). The holes were dry etched through the p+ Si layer using a SAMCO RIE-10NR Reactive Ion Etching (RIE) System. After removing the photoresist, the patterned SOI sample was dipped inside 49% HF for 3.5 hours to remove the BOX layer. The sample was then transferred to IPA solution, followed by transferring into DI water for 1 min each to completely rinse off HF. After drying up the DI water with $N_2$ air gun blow, the p+ Si membrane is ready for transfer printing for the Si/ β-$Ga_2O_3$ p-n diode fabrication (Figure 10(b)).

*Fabrication of Si/β-$Ga_2O_3$ heterostructure p-n diodes*

The p-n diode fabrication process flow was depicted in Figure 10. The grafting-ready sample (Figure 10(c)) was metallized to form a cathode contact on the backside by depositing Ti/Au/Cu/Au (10/5/100/10 nm) metal stack using e-beam evaporation followed by rapid thermal annealing (RTA) at 600 °C for 10 s to form ohmic contacts (Figure 10(d)).

The p+ Si membrane sitting on the Si handling substrate was picked up using a polydimethylsiloxane (PDMS) stamp of a thickness of ~ 10 mm. The stamp was fabricated from a mixture of monomer and crosslinker of ratio 4:1 and cured under 60 °C for 4 hours after a quick degassing process (Figure 10(e)). The p+ Si membrane was then transferred to the β-$Ga_2O_3$ substrate [62, 63]. The Si membrane transfer and printing process was performed in the ambient environment because oxygen atoms are considered beneficial to the desired double-side passivation as mentioned above. Due to the superior compliance of the Si membrane, an intimate contact can form without air trapped between the Si and β-$Ga_2O_3$ epi layer. Therefore, van der Waals bonding between Si membrane and the β-$Ga_2O_3$ epi substrate can be uniformly created. Then the weak van der Waals bonding was converted into strong chemical bonding by RTA at 350 °C for 5 min in a $N_2$ environment (Figure 10(f)). XRD and Raman spectroscopy measurements were carried out on this Si/ β-$Ga_2O_3$ stack, by a Malvern Panalytical Empyrean X-ray diffractometer with a Cu K-α X-ray source and a Horiba LabRAM HR Evolution Raman spectrometer using a 532-nm laser respectively.

Another conventional photolithography was used to cast anode contact pattern on the transferred p+ Si layer, and then an e-beam deposition of Ni/Au/Cu/Au (10/5/100/10 nm) metal stack was followed to perform lift-off process (Figure 10(g)). The individual Si/$Ga_2O_3$ pn diodes were isolated from each other by etching away the exposed Si through RIE (Figure 10(h)), with the anode metal as the self-aligned etching masks. The anode area for each p-n diode after isolation is $37 \times 37$ $\mu m^2$. A thermal annealing of 600 °C for 10 seconds was performed on the isolated Si/$Ga_2O_3$ pn diode array to improve the ohmic contacts of both anodes and cathodes. For capacitance-voltage (C-V) voltage characterizations, larger anode area Si/β-$Ga_2O_3$ p-n diodes of round shape with an area of $2.79 \times 10^4$ $\mu m^2$ were fabricated using the same fabrication process. The larger device area is to increase the diodes junction capacitance and to reduce the relative uncertainty of the capacitance measurements.

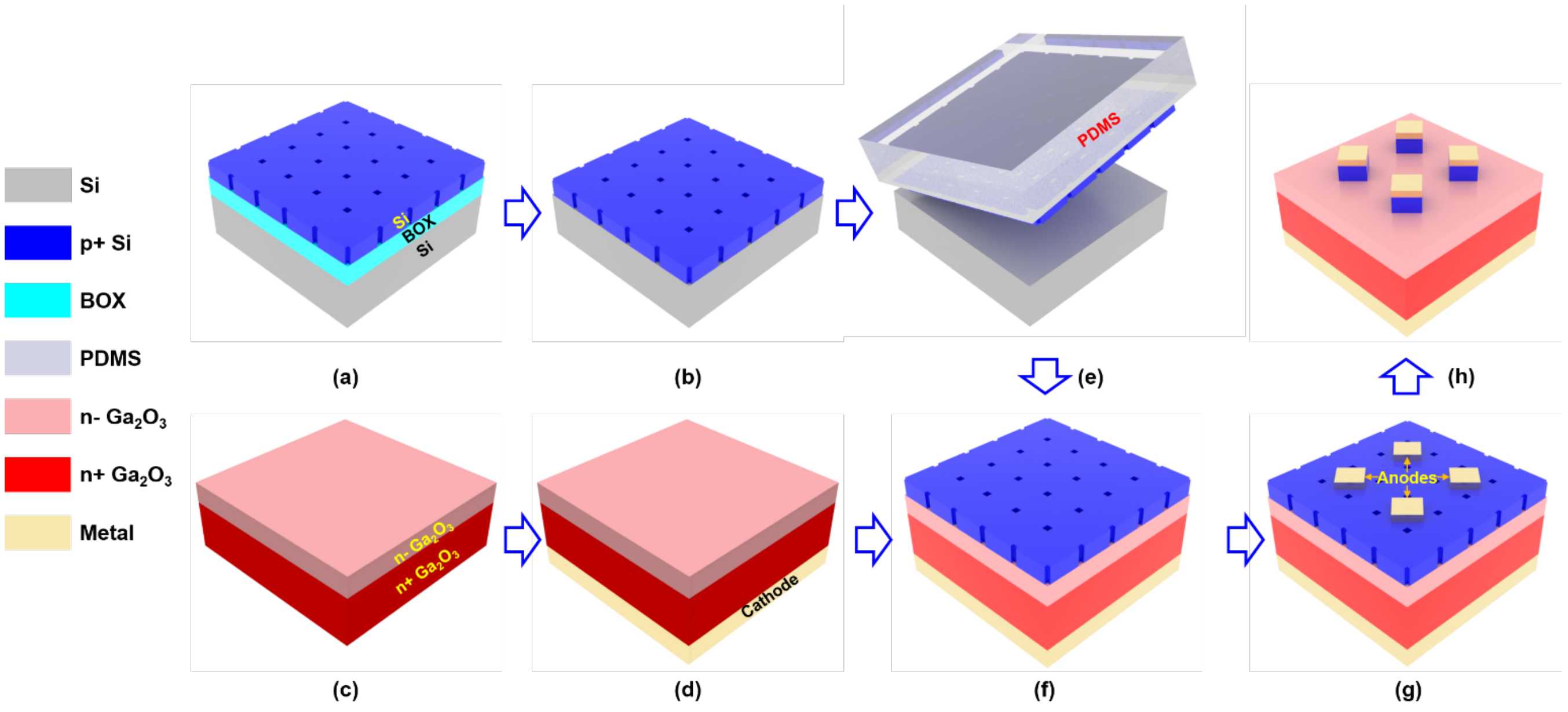

**Figure 10.** Illustration of fabrication process flow of the Si/$Ga_2O_3$ p-n diode. (a) RIE etching release holes on p+ Si device layer. (b) Release the p+ Si device layer using HF. (c) Surface preparation of the $Ga_2O_3$ epi. (d) Bottom cathode formation on the $Ga_2O_3$ epi. (e) Pick up of p+ Si membrane using a PDMS stamp. (f) Transfer-printing of the p+ Si membrane to the receiver $Ga_2O_3$ substrate. (g) Anode contact formation on the p+ Si layer. (h) Si/$Ga_2O_3$ pn diodes isolation and contacts ohmic thermal annealing.

*I-V and C-V characterizations*

The I-V and C-V of the Si/β-$Ga_2O_3$ p-n diodes were measured by a Keithley 4200 semiconductor characterization system. The device reverse breakdown characteristics were measured using HP 4155B semiconductor parameter analyzer.

*Si/β-$Ga_2O_3$ interface imaging*

The transmission electron microscopy (TEM) lamella, which was taken from an I-V characterized and functional Si/$Ga_2O_3$ pn diode, was prepared using a Thermo Fisher Scientific (TFS) Helios G4 dual-beam focused ion beam (FIB) system equipped with an EasyLift manipulator for the transfer of the lamella to the Omniprobe FIB grid. To protect the surface morphology of the device, the electron beam-assisted carbon (C) and Pt deposition, followed by the Ga ion beam assisted Pt deposition were performed prior to the Ga ion milling. 30 kV Ga ion milling was used for the bulk milling, followed by 5 kV and 2 kV Ga ion milling, *i.e.*, low kV cleaning, to remove the damaged layer created by the 30 kV Ga ion milling.

Scanning transmission electron microscopy (STEM) micrographs were acquired using a Thermo Fisher Scientific (TFS) Themis Z aberration-corrected electron microscope operated at 300 keV. The convergence semi-angle of 21 mrad and the collection semi-angle of 70-200 mrad were used for the high angle annular dark-field (HAADF) image acquisition. Electron energy loss spectroscopy (EELS) analysis was performed using a Gatan Continuum electron energy loss spectrometer and image filter and K3 electron direct detector. The collection semi-angle of 50 mrad was used to acquire the EELS spectra.

**Acknowledgements**

The work was supported by a CRG grant (2022-CRG11-5079.2) by the King Abdullah University of Science and Technology (KAUST).

**Conflict of Interest**

The authors declare no conflict of interest.

**Data Availability Statement**

The data that support the findings of this study are available from the corresponding authors upon reasonable request.

semiconductor grafting, Gallium oxide, heterojunction, diodes, scanning transmission electron microscopy, monocrystalline membrane, transfer-printing, ultra-wide bandgap (UWBG), X-ray photoelectron spectroscopy, Raman spectroscopy

The preliminary results contained in the manuscript was presented as a poster in

**The 5th US Gallium Oxide Workshop**

Washington, D.C. USA

August 07-10, 2022

# Grafted Si/$Ga_2O_3$ p-n diodes

Hokyung Jang[1], Donghyeok Kim[1], Jiarui Gong[1], Fikadu Alema[2], Andrei Osinsky[2], Kelson Chabak[3], Gregg Jessen[3], Vincent Gambin[4], Shubhra S Pasayat[1], Chirag Gupta[1], Zhenqiang Ma[1,a]
(1)University of Wisconsin-Madison, (2) Agnitron Technology Inc., (3) Air Force Research Laboratory, (4)Northrop Grumman
(a)Tel/Fax: +1 (608) 261-1095; E-mail: mazq@engr.wisc.edu

## 1. Introduction

**(1) Why $Ga_2O_3$?**

- High breakdown electrical field (8 MV/cm).
- High electron saturation velocity (1.1 × $10^7$ cm/s).
- Ultra-Wide Bandgap: 4.9 eV
- Available up to 4-inch wafers.

**(2) Why no $Ga_2O_3$ bipolar devices? → Ineffective p-type doping in $Ga_2O_3$.**

Doping effectives among semiconductors

| Material | Eg (eV) | $\mu_e$ ($cm^2$/V·s) | $\mu_h$ ($cm^2$/V·s) | $V_{sat}$ ($10^7$ cm/s) | $E_{br}$ (MV/cm) | $\sigma_{th}$ (W/cm·K) | N-type | P-type |
|---|---|---|---|---|---|---|---|---|
| $Ga_2O_3$ | 4.9 | 180 | - | 1.1 | 8 | 0.11-0.27 | Good | Poor |
| Si | 1.12 | 1350 | 450 | 1 | 0.3 | 1.5 | Good | Good |
| Ge | 0.66 | 3900 | 1900 | 0.6 | 0.1 | 0.58 | Good | Good |
| GaAs | 1.42 | 8500 | 400 | 1 | 0.4 | 0.43 | Good | Good |
| InP | 1.34 | 5400 | 200 | 0.67 | 0.5 | 0.68 | Good | Good |

**(3) Towards Lattice-Mismatched Heterojunctions:**
**Semiconductor Grafting**

**What is GRAFTING technique in nature?**

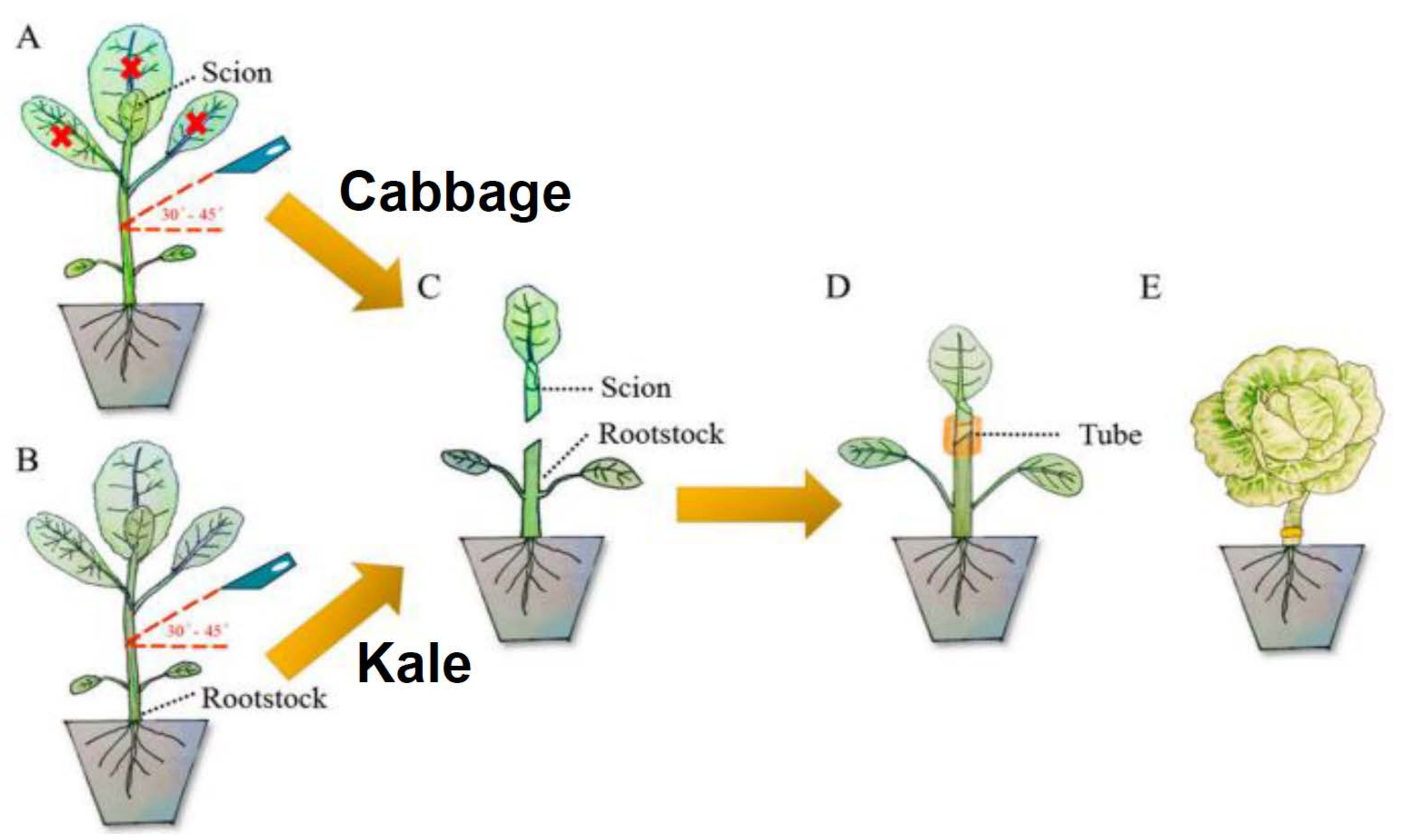

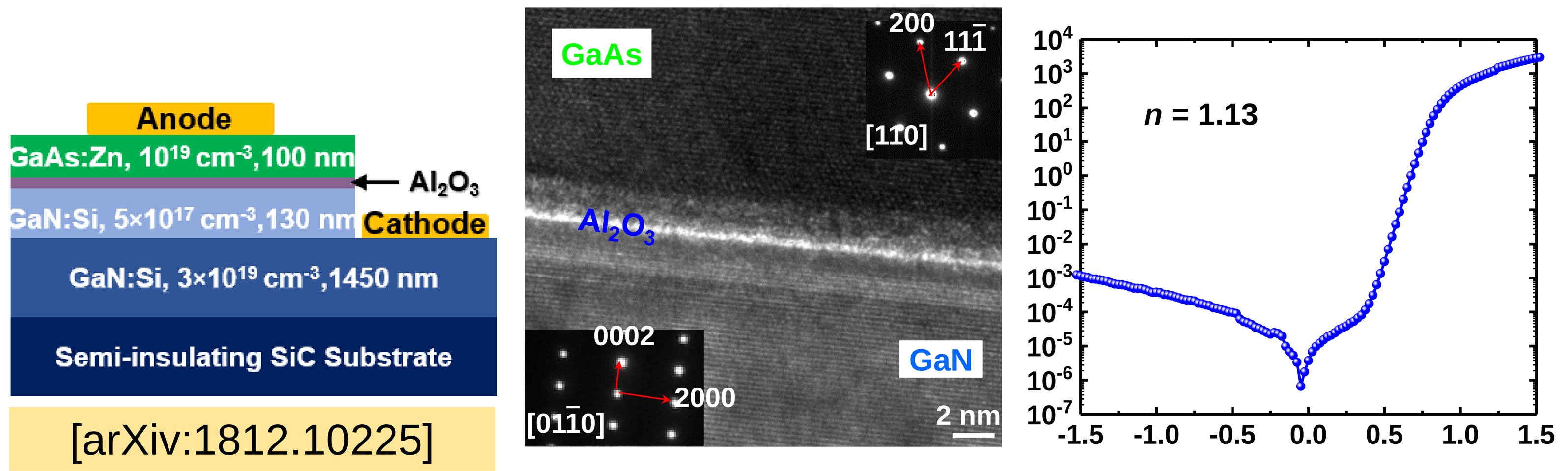

- Semiconductor grafting provided an effective solution to lattice-mismatch encountered by epitaxial growth community.
- **Ultrathin oxide** functions as a passivation and quantum tunneling layer.
- Ultra-low density of interface states ($D_{it}$): **$10^{10}$-$10^{11}$ /$cm^3$.eV**

## 2. Experiment

**Schematic device layer structure**

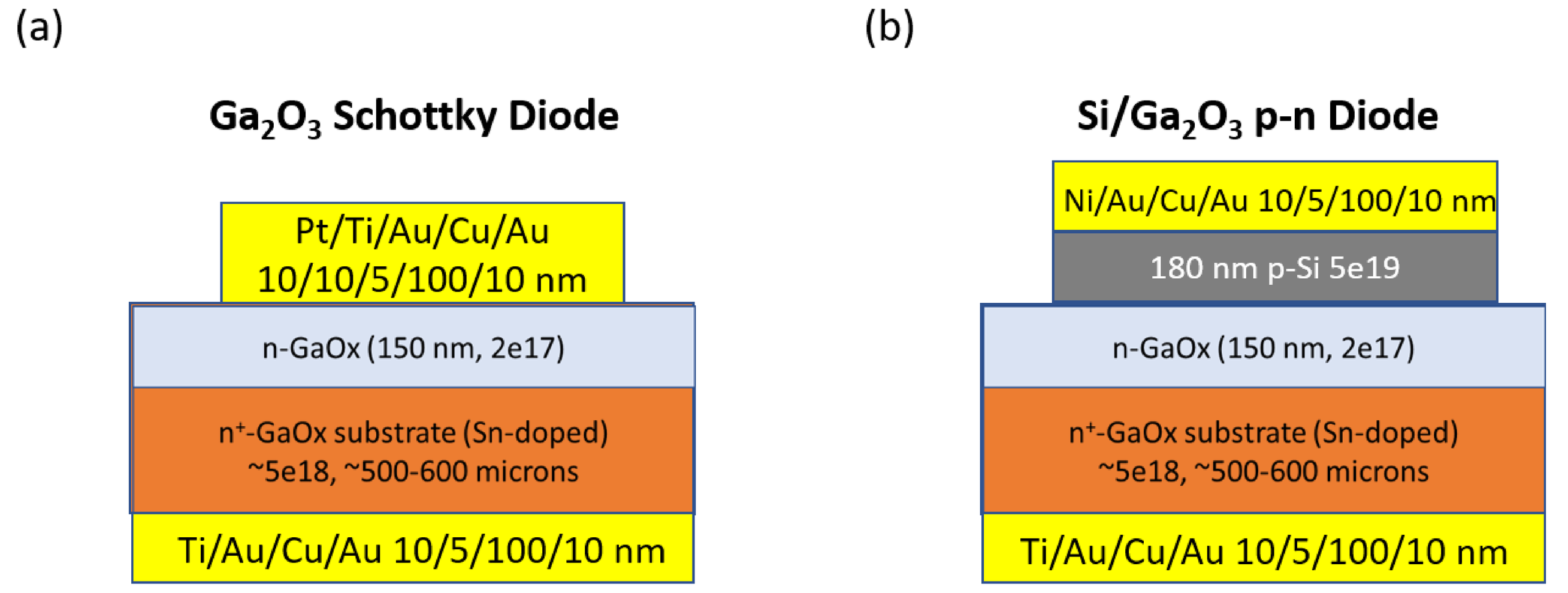

**Schematic fabrication process flow**

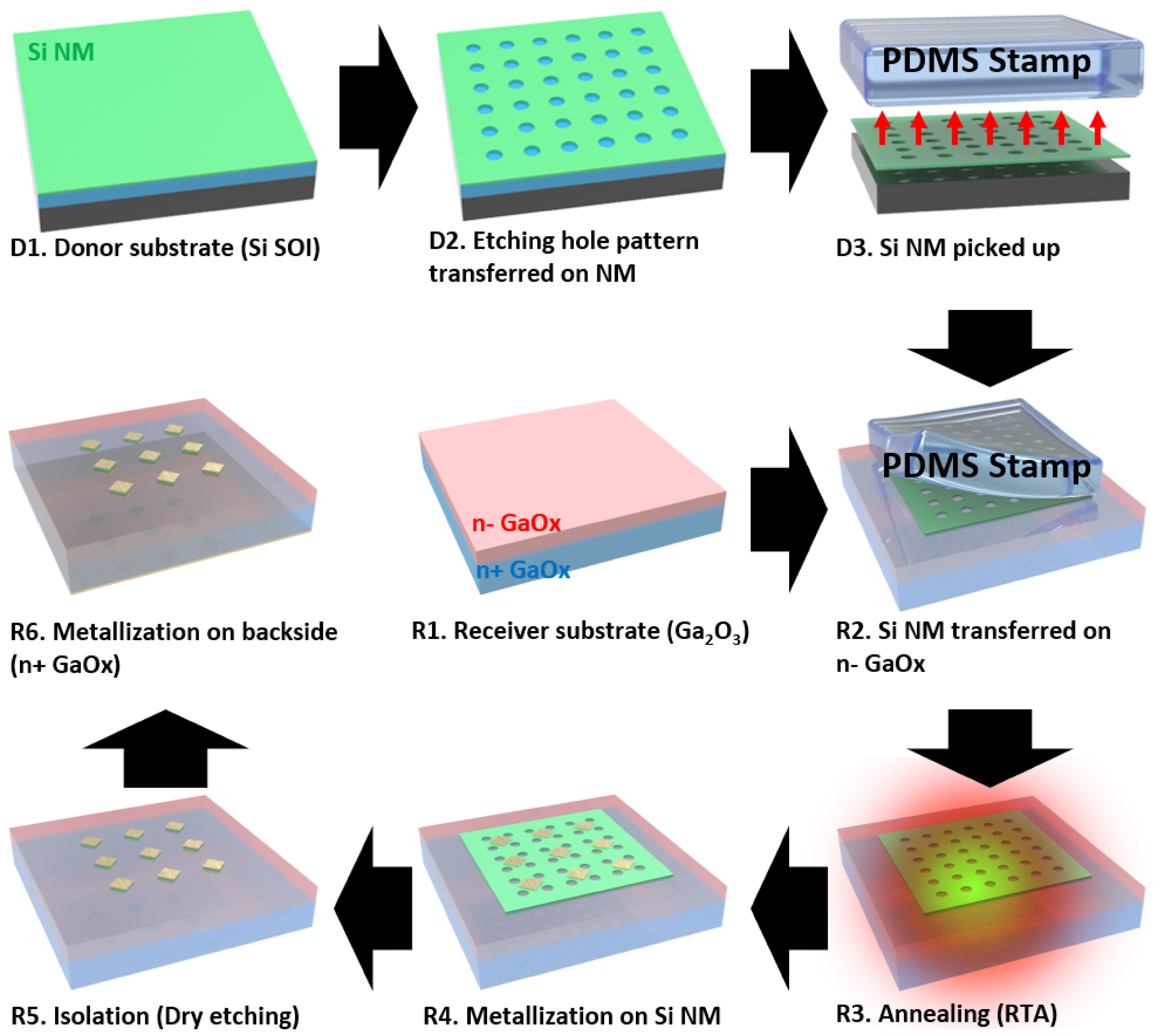

**Optical images of the device array**

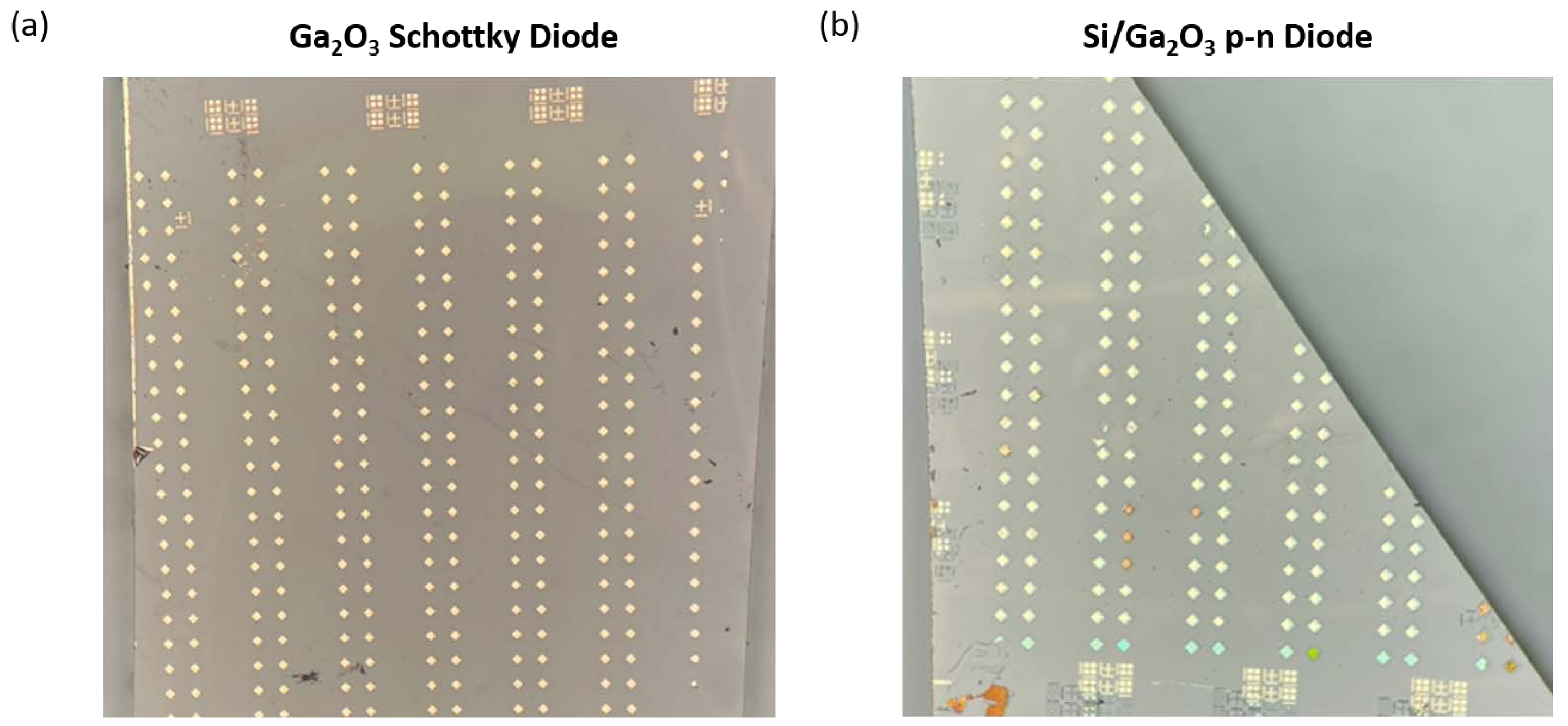

## 3. Characterizations:

**(1) IV measurement**

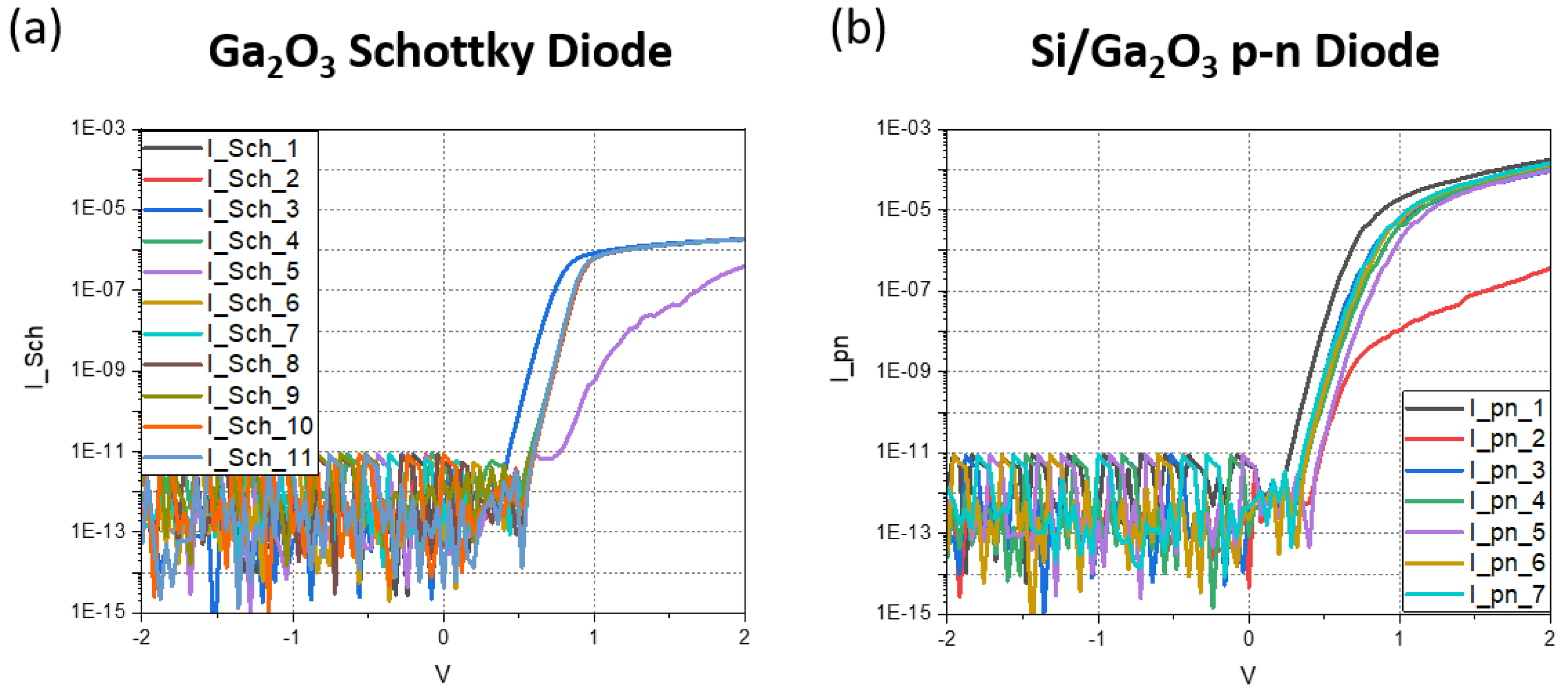

Ideality factor n=1.10
Current density I=118 mA/$cm^2$ @2V

Ideality factor n=1.17
Current density I=1094 mA/$cm^2$ @2V

**(2) Breakdown measurement**

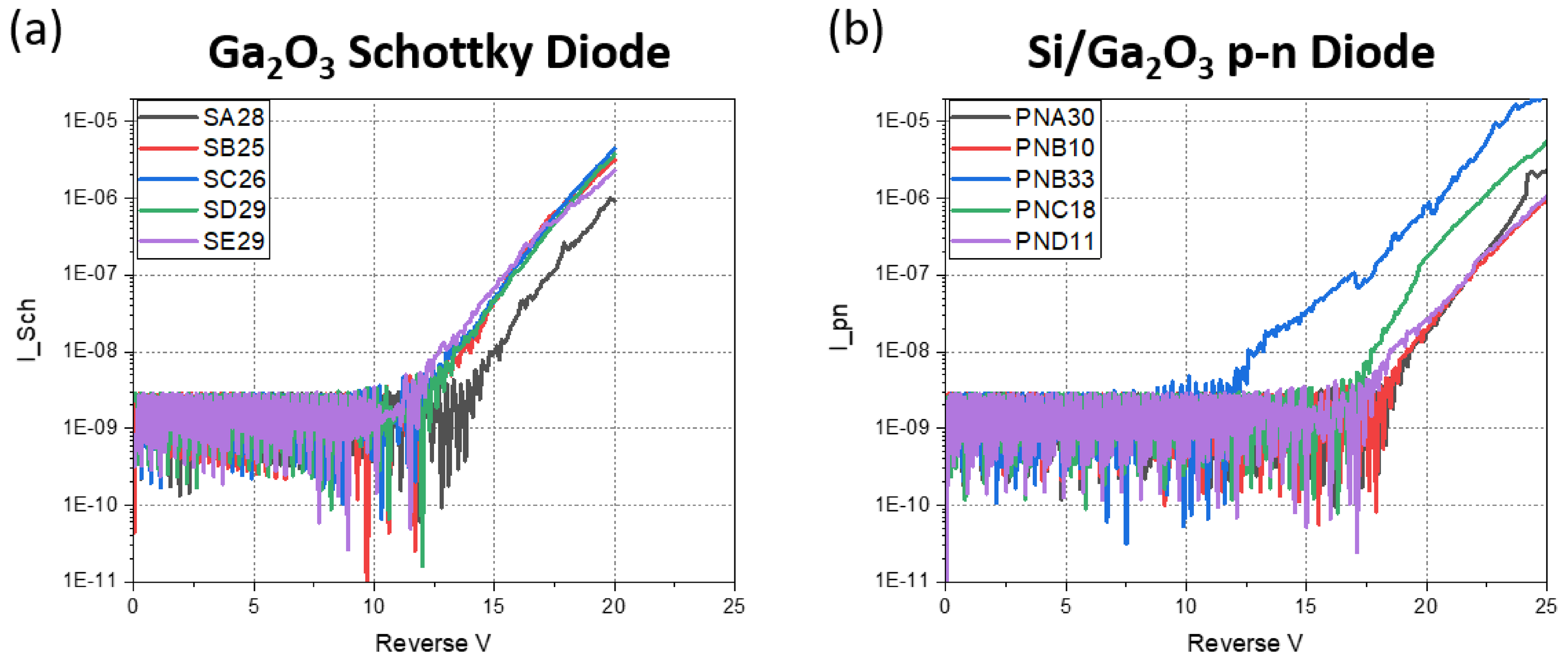

**$Ga_2O_3$ Schottky**

| Device # | A28 | B25 | C26 | D29 | E29 |
|---|---|---|---|---|---|
| BD (V) | 20 | 18.3 | 18.1 | 18.4 | 18.6 |

**Si/$Ga_2O_3$ (Annealed)**

| Device # | A30 | B10 | B33 | C18 | D11 |
|---|---|---|---|---|---|
| BD (V) | 24.2 | 25 | 20.7 | 22.4 | 24.9 |

## 4. Conclusions:

- High-quality grafted Si/$Ga_2O_3$ p-n diodes were fabricated and characterized.
- Higher on/off ratio, good ideality factor, and larger breakdown voltage have been confirmed on the grafted devices than the Schottky diodes.
- Bipolar transistors become feasible.

# GOX 2022 Program Overview

| Room /Time | Jefferson 1 & Atrium | Jefferson 2-3 |
|---|---|---|
| MoM | | KEY1: Keynote Address<br>AC-MoM: Characterization & Modeling I<br>BG-MoM: Bulk & Epitaxy I |
| MoA | | MD-MoA: Process & Devices I<br>TM-MoA: Characterization & Modelling II |
| MoP | Poster Sessions:<br>Advanced Characterization Techniques (AC)<br>Dielectric Interfaces (DI)<br>Electronic and Photonic Devices, Circuits and Applications (EP)<br>Electronic Transport &Breakdown Phenomena (ET)<br>Heterogeneous Material Integration (HM) | |
| TuM | | AC-TuM: Advanced Characterization & Microscopy<br>PS1-TuM: Plenary Session I<br>TM-TuM: Characterization & Modelling III |
| TuA | | DI-TuA: Processes & Devices II<br>EG-TuA: Bulk & Epitaxy II |
| TuP | Poster Sessions:<br>Epitaxial Growth (EG)<br>Material and Device Processing & Fabrication Techniques (MD)<br>Theory, Modeling and Simulation (TM) | |
| WeM | | PS2-WeM: Plenary Session II<br>EP1-WeM: Process & Devices III<br>EP2-WeM: Process and Devices IV |

# Monday Morning, August 8, 2022

## Room Jefferson 2-3

| Time | Presentation | Session |
|---|---|---|
| 8:30am | **Welcome and Sponsor Thank Yous** | **Keynote Address**<br>**Session KEY1**<br>**Keynote Address**<br>**Moderator:**<br>**Kelson Chabak**, Air Force Research Laboratory |
| 8:45am | **INVITED: KEY1-2** Keynote Lecture: $Ga_2O_3$ Device Technologies: Power Switching and High-Frequency Applications, and Beyond, ***Masataka Higashiwaki***, Department of Physics and Electronics, Osaka Metropolitan University, Japan; *T. Kamimura, S. Kumar, Z. Wang,* National Institute of Information and Communications Technology, Japan; *T. Kitada, J. Liang, N. Shigekawa,* Department of Physics and Electronics, Osaka Metropolitan University, Japan; *H. Murakami, Y. Kumagai,* Department of Applied Chemistry, Tokyo University of Agriculture and Technology, Japan | |
| 9:30am | **INVITED: AC-MoM-5** Characterization of Deep Acceptors in β-$Ga_2O_3$ by Deep Level Optical Spectroscopy, *H. Ghadi, J. McGlone, E. Cornuelle,* The Ohio State University; *A. Senckowski,* University of Massachusetts Lowell; *S. Sharma, U. Singisetti,* University of Buffalo; *M. Wong,* University of Massachusetts Lowell; *A. Arehart,* ***Steven A Ringel***, The Ohio State University | **Advanced Characterization Techniques**<br>**Session AC-MoM**<br>**Characterization & Modeling I**<br>**Moderator:**<br>**Kornelius Tetzner**, Ferdinand-Braun-Institut, Leibniz-Institut für Höchstfrequenztechnik (FBH), Germany |
| 10:00am | **AC-MoM-7** Determination of Cation Vacancy and Al Diffusion Constants in B-$(Al,Ga)_2O_3$ / $Ga_2O_3$ Superlattices, *H. Yang, A. Levin, B. Eisner, A. Bhattacharyya, P. Ranga, S. Krishnamoorthy,* ***Michael Scarpulla***, University of Utah | |
| 10:15am | **AC-MoM-8** Defect Characterization in Gallium Oxide and Related Materials Using Terahertz Electron Paramagnetic Resonance Ellipsometry: Fe in $Ga_2O_3$, ***Mathias Schubert***, University of Nebraska, Lincoln; *S. Richter,* Lund University, Sweden; *S. Knight, P. Kuehne,* Linkoping University, Sweden; *M. Stokey, R. Korlacki,* University of Nebraska-Lincoln; *V. Stanishev,* Linkoping University, Sweden; *Z. Galazka, K. Irmscher,* Leibniz-Institut fuer Kristallzuechtung, Germany; *S. Mu, C. Van de Walle,* University of California at Santa Barbara; *V. Ivády,* MPI Physics of Complex Systems, Germany; *O. Bulancea-Lindvall, I. Abrikosov,* Linkoping University, Sweden; *V. Darakchieva,* Lund University, Sweden | |
| 10:30am | **BREAK** | |
| 10:45am | **INVITED: BG-MoM-10** b-$Ga_2O_3$ Growth and Wafer Fabrication, *A. Brady, G. Foundos,* ***Chase Scott***, Northrop Grumman SYNOPTICS; *V. Gambin,* Northrop Grumman Corporation; *K. Stevens,* Northrop Grumman SYNOPTICS; *J. Blevins,* Air Force Research Laboratory, Afghanistan | **Bulk Growth**<br>**Session BG-MoM**<br>**Bulk & Epitaxy I**<br>**Moderator:**<br>**John Blevins**, Air Force Research Laboratory |
| 11:15am | **BG-MoM-12** Increasing the Bandgap of β-$Ga_2O_3$ via Alloying with $Al_2O_3$ or $Sc_2O_3$ in Czochralski-grown Crystals, ***Benjamin Dutton***, *J. Jesenovec, B. Downing, J. McCloy,* Washington State University | |
| 11:30am | **BG-MoM-13** Chemi-Mechanical Polishing and Subsurface Damage Characterization of 2-inch (010) Semi-Insulating β-$Ga_2O_3$ Substrates, ***David Snyder***, Penn State Applied Research Laboratory | |
| 11:45am | **BG-MoM-14** Ge-Delta Doped *β*-$Ga_2O_3$ Grown Via Plasma Assisted Molecular Beam Epitaxy, ***Thaddeus Asel***, Air Force Research Laboratory, Materials and Manufacturing Directorate, USA; *E. Steinbrunner,* Wright State University, Department of Electrical Engineering; *J. Hendrick,* Air Force Institute of Technology, Department of Engineering Physics; *A. Neal, S. Mou,* Air Force Research Laboratory, Materials and Manufacturing Directorate, USA | |
| 12:00pm | **BG-MoM-15** High Purity n-type β-$Ga_2O_3$ Films with $10^{13}$ $cm^{-3}$ Residual Accepter Concentration by MOCVD, ***Andrei Osinsky***, *F. Alema,* Agnitron Technology | |

# Monday Afternoon, August 8, 2022

## Room Jefferson 2-3

### Material and Device Processing and Fabrication Techniques
### Session MD-MoA
### Process & Devices I
**Moderator:**
**Man-Hoi Wong**, University of Massachusetts Lowell

| Time | Presentation |
| --- | --- |
| 1:45pm | **INVITED: MD-MoA-1** High Aspect Ratio $Ga_2O_3$-based Homo and H0eterostructures by Plasma-free Metal-assisted Chemical Etching, ***Xiuling Li***, University of Texas at Austin; *H. Huang, C. Chan, J. Michaels,* University of Illinois, Urbana-Champaign |
| 2:15pm | **MD-MoA-3** Blocking Behavior of N and Fe Ion Implanted β-$Ga_2O_3$, ***Bennett Cromer***, Cornell University; *W. Li,* University of California at Berkeley; *K. Smith,* Cornell University; *K. Gann,* Cornell University, Iceland; *K. Nomoto,* Cornell University; *N. Hendriks,* University of California at Santa Barbara; *A. Green, K. Chabak,* Air Force Research Laboratory; *M. Thompson, D. Jena, G. Xing,* Cornell University |
| 2:30pm | **MD-MoA-4** Evolution and Recovery of Ion Implantation-Induced Damage Zone in β-Ga2O3, ***Elaf Anber***, *D. Foley, J. Nathaniel,* Johns Hopkins University; *A. Lang ,* American Society for Engineering Education; *J. Hart ,* Johns Hopkins University; *M. Tadjer , K. Hobart,* US Naval Research Laboratory; *S. Pearton,* University of Florida, Gainesville; *M. Taheri,* Johns Hopkins University |
| 2:45pm | **MD-MoA-5** Heterogeneous Integration of Single-Crystal β-$Ga_2O_3$ and N-Polar GaN Substrates With ZnO Interlayer Deposited by Atomic Layer Deposition, ***Zhe (Ashley) Jian***, University of Michigan, Ann Arbor; *C. Clymore,* University of California, Santa Barbara; *D. Agapiou,* University of Michigan, Ann Arbor; *U. Mishra,* University of California, Santa Barbara; *E. Ahmadi,* University of Michigan, Ann Arbor |
| 3:00pm | **MD-MoA-6** Structural Transformation of β-$Ga_2O_3$ through Si-implantation, ***Snorre Braathen Kjeldby***, *A. Azarov, P. Nguyen,* Centre for Materials Science and Nanotechnology, University of Oslo, Norway; *V. Venkatachalapathy,* Centre for Materials Science and Nanotechnology, University of Oslo and Department of Materials Science, National Research Nuclear University, "MEPhI", Norway; *R. Mikšová,* Nuclear Physics Institute of the Czech Academy of Sciences, Czechia; *A. Macková,* Nuclear Physics Institute of the Czech Academy of Sciences and Department of Physics, Faculty of Science, J.E. Purkyně University, Czechia; *J. García-Fernández, A. Kuznetsov, Ø. Prytz, L. Vines,* Centre for Materials Science and Nanotechnology, University of Oslo, Norway |
| 3:15pm | **MD-MoA-7** Electrical Characteristics of *in Situ* Mg-Doped $Ga_2O_3$ Current-Blocking Layer for Vertical Devices, ***Sudipto Saha***, University at Buffalo-SUNY; *L. Meng, A. Bhuiyan, Z. Feng, H. Zhao,* Ohio State University; *U. Singisetti,* University at Buffalo-SUNY |
| 3:30pm | **BREAK** |

### Theory, Modeling and Simulation
### Session TM-MoA
### Characterization & Modelling II
**Moderator:**
**Mike Thompson**, Cornell University

| Time | Presentation |
| --- | --- |
| 3:45pm | **INVITED: TM-MoA-9** Transport, Doping, and Defects in β-$Ga_2O_3$, ***Adam Neal***, Air Force Research Laboratory, Materials and Manufacturing Directorate, USA |
| 4:15pm | **TM-MoA-11** Structural Changes to Beta Gallium Oxide from Ion Irradiation Damage: Model and Relation to in-Situ Experiments, ***Alexander Petkov***, *D. Cherns, D. Liu,* University of Bristol, UK; *W. Chen, M. Li,* Argonne National Laboratory, USA; *J. Blevins,* Air Force Research Laboratory, USA; *V. Gambin,* Northrop Grumman; *M. Kuball,* University of Bristol, UK |
| 4:30pm | **TM-MoA-12** Band Structure Across κ-$(In_xGa_{1-x})_2O_3$/κ-$(Al_yGa_{1-y})_2O_3$ Thin Film Interfaces, ***Ingvild Julie Thue Jensen***, *A. Thøgersen, E. Fertitta, B. Belle,* SINTEF Materials Physics, Norway; *A. Langørgen, S. Cooil, Y. Hommedal, Ø. Prytz, J. Wells, L. Vines,* University of Oslo, Norway; *H. von Wenckstern,* University of Leipzig, Germany |
| 4:45pm | **TM-MoA-13** Aluminum Incorporation Striations in (-201) β-$(Al_xGa_{1-x})_2O_3$ Films Grown on C-Plane and Miscut Sapphire Substrates, ***Kenny Huynh***, *Y. Wang, M. Liao,* University of California Los Angeles; *P. Ranga,* University of Utah; *S. Krishnamoorthy,* University of California at Santa Barbara; *M. Goorsky,* University of California, Los Angeles |
| 5:00pm | **TM-MoA-14** Plasmon-phonon Coupling in Electrostatically Gated *β*-$Ga_2O_3$ Films with Mobility Exceeding 200 $cm^2V^{-1}s^{-1}$, *A. Rajapitamahuni, A. Manjeshwar,* University of Minnesota, USA; *A. Kumar, A. Datta,* University at Buffalo; *P. Ranga,* University of California Santa Barbara; *L. Thoutam,* SR University, Warangal, India; *S. Krishnamoorthy,* University of California Santa Barbara; ***Uttam Singisetti***, University at Buffalo; *B. Jalan,* University of Minnesota, USA |

## Advanced Characterization Techniques
## Room Jefferson 1 & Atrium - Session AC-MoP
## Advanced Characterization Techniques Poster Session
## 5:15pm – 7:15pm

**AC-MoP-1** Advanced Defect Characterization in b-$Ga_2O_3$ Without the Arrhenius Plot, ***Jian Li,*** NCKU, Taiwan*; A. Neal, S. Mou,* Air Force Research Laboratory, Materials and Manufacturing Directorate, USA*; M. Wong,* University of Massachusetts Lowell

**AC-MoP-2** Infrared-Active Phonon Modes and Static Dielectric Constants of Orthorhombic $LiGaO_2$, ***Teresa Gramer****, M. Stokey, R. Korlacki, M. Schubert,* University of Nebraska - Lincoln

**AC-MoP-3** Spectroscopic Ellipsometry Optical Analysis of Zinc Gallate at Elevated Temperatures, ***Emma Williams****,* University of Nebraska-Lincoln, USA*; M. Hilfiker, U. Kilic, Y. Traouli, N. Koeppe, J. Rivera, A. Abakar, M. Stokey, R. Korlacki,* University of Nebraska - Lincoln*; Z. Galazka,* Leibniz-Institut für Kristallzüchtung, Germany*; M. Schubert,* University of Nebraska - Lincoln

**AC-MoP-4** The Electron Spin Hamiltonian for $Fe^{3+}$ in Monoclinic β-$Ga_2O_3$, ***Steffen Richter****,* Lund University, Sweden*; S. Knight, P. Kühne,* Linköping University, Sweden*; M. Schubert,* University of Nebraska - Lincoln*; V. Darakchieva,* Lund University, Sweden

**AC-MoP-5** Characterization of (010) β-$Ga_2O_3$ to Support Fabrication, Wafer Size Scaleup, and Epi Development, ***David Snyder****,* Penn State Applied Research Laboratory

**AC-MoP-6** Photoluminescence Spectroscopy of $Cr^{3+}$ in β-$Ga_2O_3$ and $(Al_{0.1}Ga_{0.9})_2O_3$ , ***Cassandra Remple****, J. Jesenovec, B. Dutton, J. McCloy, M. McCluskey,* Washington State University

**AC-MoP-7** Surface Relaxation and Rumpling of Sn Doped B-ga2o3(010), ***Nick Barrett****,* CEA Saclay, France*; A. Pancotti,* Universidade Federal de Jataí, Brazil*; T. Back,* AFRL*; W. Hamouda, M. Laccheb, C. Lubin, A. Boucly,* CEA Saclay, France*; P. Soukiassian,* Université Paris-Saclay, France*; J. Boeckl, D. Dorsey, S. Mou, T. Asel,* AFRL*; G. Geneste,* CEA, France

**AC-MoP-8** Probing Vacancies and Hydrogen Related Defects in β-$Ga_2O_3$ with Positrons and FTIR, ***Corey Halverson****, M. Weber, J. Jesenovec, B. Dutton, C. Remple, M. McCluskey, J. McCloy,* Washington State University

**AC-MoP-9** Evolution of Anisotropy and Order of Band-to-Band Transitions, Excitons, Phonons, Static and High Frequency Dielectric Constants Including Strain Dependencies in Alpha and Beta Phase $(Al_xGa_{1-x})_2O_3$ , ***Megan Stokey,*** University of Nebraska-Lincoln*; R. Korlacki, M. Hilfiker, T. Gramer,* University of Nebraska - Lincoln*; J. Knudtson,* University of Nebraska-Lincoln*; S. Richter,* Lund University, Sweden*; S. Knight,* Linköping University, Sweden*; A. Mock,* Weber State University*; A. Mauze, Y. Zhang, J. Speck,* University of California Santa Barbara*; R. Jinno, Y. Cho, H. Xing, D. Jena,* Cornell University*; Y. Oshima,* National Institute for Materials Science, Japan*; E. Ahmadi,* University of Michigan*; V. Darakchieva,* Lund University, Sweden*; M. Schubert,* University of Nebraska - Lincoln

**AC-MoP-10** Photoluminescence Mapping of Gallium Oxide and Aluminum Gallium Oxide Epitaxial Films, ***Jacqueline Cooke****, P. Ranga,* University of Utah*; J. Jesenovec, J. McCloy,* Washington State University*; S. Krishnamoorthy,* University of California at Santa Barbara*; M. Scarpulla, B. Sensale-Rodriguez,* University of Utah

**AC-MoP-11** Cathodoluminescence (CL) Evaluation of Silicon Implant Activation and Damage Annealing in Beta Ga2O3 EPI in Heavily Silicon Doped Contact Regions, ***Stephen Tetlak****,* Air Force Research Laboratory*; K. Gann, J. McCandless,* Cornell University*; K. Liddy,* Air Force Research Laboratory*; D. Jenna, M. Thompson,* Cornell University

**AC-MoP-12** Non-Destructive Characterization of Annealed Si-Implanted Thin Film β-$Ga_2O_3$, ***Aine Connolly****, K. Gann,* Cornell University*; S. Tetlak,* Air Force Research Laboratory*; V. Protasenko,* Cornell University*; M. Slocum, S. Mou,* Air Force Research Laboratory*; M. Thompson,* Cornell University

## Dielectric Interfaces
## Room Jefferson 1 & Atrium - Session DI-MoP
## Dielectric Interfaces Poster Session
## 5:15pm – 7:15pm

**DI-MoP-1** Band Offsets of MOCVD Grown β-(Al0.21Ga0.79)2O3/β-Ga2O3 (010) Heterojunctions*, T. Morgan, J. Rudie, M. Zamani-Alavijeh, A. Kuchuk,* University of Arkansas*; N. Orishchin, F. Alema,* Agnitron Technology Incorporated*; A. Osinsky,* Agnitron Technology Incorporated, United States Minor Outlying Islands (the)*; R. Sleezer,* Minnesota State University at Mankato*; G. Salamo,* University of Arkansas, United States Minor Outlying Islands (the)*; **Morgan Ware**,* University of Arkansas

**DI-MoP-2** Optimization of MOCVD Grown In-situ Dielectrics for β-$Ga_2O_3$*, G. Wang,* University of Wisconsin - Madison*; F. Alema,* Agnitron Technology Inc.*; J. Chen,* University of Wisconsin - Madison*; A. Osinsky,* Agnitron Technology Inc.*; C. Gupta,* University of Wisconsin-Madison*; **Shubhra Pasayat**,* University of Wisconsin - Madison

## Electronic and Photonic Devices, Circuits and Applications
## Room Jefferson 1 & Atrium - Session EP-MoP
## Electronic and Photonic Devices, Circuits and Applications Poster Session – 5:15pm – 7:15pm

**EP-MoP-1** Investigating Ohmic Contacts for High Temperature Operation of $Ga_2O_3$ Devices*, **William Callahan**,* Colorado School of Mines*; S. Sohel,* National Renewable Energy Laboratory*; M. Sanders, R. O'Hayre,* Colorado School of Mines*; D. Ginley, A. Zakutayev,* National Renewable Energy Laboratory

**EP-MoP-2** Gate Effects of Channel and Sheet Resistance in β-$Ga_2O_3$ Field-Effect Transistors using the TLM Method*, **Ory Maimon**,* Department of Electrical Engineering, George Mason University*; N. Moser,* Air Force Research Laboratory, Sensors Directorate*; K. Liddy, A. Green, K. Chabak,* Air Force Research Laboratory, Sensors Directorate, USA*; C. Richter, K. Cheung, S. Pookpanratana,* Nanoscale Device and Characterization Division, National Institute of Standards and Technology*; Q. Li,* Department of Electrical Engineering, George Mason University

**EP-MoP-3** Lateral β-$Ga_2O_3$ Schottky Barrier Diodes With Interdigitated Contacts*, **Jeremiah Williams**,* Air Force Research Laboratory, Sensors Directorate*; A. Arias-Purdue,* Teledyne*; K. Liddy, A. Green,* Air Force Research Laboratory, Sensors Directorate*; D. Dryden, N. Sepelak,* KBR*; K. Singh,* Air Force Research Laboratory, Sensors Directorate*; F. Alema, A. Osinsky,* Agnitron Technology*; A. Islam, N. Moser, K. Chabak,* Air Force Research Laboratory, Sensors Directorate

**EP-MoP-4** Optimized Annealing for Activation of Implanted Si in β-Ga2O3*, **Katie Gann**, J. McCandless,* Cornell University*; T. Asel, S. Tetlak,* Air Force Research Laboratory*; D. Jena, M. Thompson,* Cornell University

## Electronic Transport and Breakdown Phenomena
## Room Jefferson 1 & Atrium - Session ET-MoP
## Electronic Transport and Breakdown Phenomena Poster Session
## 5:15pm

**ET-MoP-2** Electric Field Mapping in β-$Ga_2O_3$ by Photocurrent Spectroscopy*, **Darpan Verma**, M. Adnan, S. Dhara,* Ohio State University*; C. Sturm,* Universitat Leipzig, Germany*; S. Rajan, R. Myers,* Ohio State University

**ET-MoP-3** Activation of Si, Ge, and Sn Donors in High-Resistivity Halide Vapor Phase Epitaxial β-$Ga_2O_3$:N*, **Joseph Spencer**,* Naval Research Laboratory/ Virginia Tech CPES*; M. Tadjer, A. Jacobs, M. Mastro, J. Gallagher, J. Freitas, Jr,* Naval Research Laboratory*; T. Tu, A. Kuramata, K. Sasaki,* Novel Crystal, Japan*; Y. Zhang,* Virginia Tech (CPES)*; T. Anderson, K. Hobart,* Naval Research Laboratory

## Heterogeneous Material Integration
## Room Jefferson 1 & Atrium - Session HM-MoP
## Heterogeneous Material Integration Poster Session
## 5:15pm – 7:15pm

**HM-MoP-1** Structural and Thermal Transport Analysis of Wafer Bonded β-$Ga_2O_3$ |4H-SiC*, **Michael Liao**, K. Huynh, Y. Wang,* UCLA*; Z. Cheng,* UIUC*; J. Shi,* GaTech*; F. Mu,* IMECAS, China*; T. You, W. Xu, X. Ou,* ShanghaiTech, China*; T. Suga,* Meisei University, Japan*; S. Graham,* GaTech*; M. Goorsky,* UCLA

**HM-MoP-2** Advances in Plasma-Enhanced Atomic Layer Deposited (PEALD) $Ga_2O_3$ Films*, **Virginia Wheeler**, A. Lang, N. Nepal, E. Jin, D. Katzer, V. Gokhale, B. Downey, D. Meyer,* US Naval Research Laboratory

**HM-MoP-3** Grafted Si/Ga2O3 pn Diodes*, H. Jang, D. Kim,* University of Wisconsin - Madison*; J. Gong,* University of Wisconsin at Madison*; F. Alema, A. Osinsky,* Agnitron Technology Inc.*; K. Chabak,* Air Force Research Laboratory*; G. Jessen,* BAE Systems*; G. Vincent,* Northrup Grumann*; S. Pasayat, C. Gupta,* University of Wisconsin - Madison*; **Zhenqiang Ma**,* 1415 Engineering Drive

# Tuesday Morning, August 9, 2022

| | Room Jefferson 2-3 | |
|---|---|---|
| 8:30am | **Welcome and Sponsor Thank Yous** | **Plenary Session**<br>**Session PS1-TuM**<br>**Plenary Session I**<br>**Moderator:**<br>**Kelson Chabak**, Air Force Research Laboratory |
| 8:45am | **INVITED: PS1-TuM-2** Plenary Lecture: Gallium Oxide Electronics - Device Engineering Toward Ultimate Material Limits, ***Siddharth Rajan***, The Ohio State University | |
| 9:15am | **INVITED: TM-TuM-4** First-Principles Modeling of $Ga_2O_3$, ***Hartwin Peelaers***, University of Kansas | **Theory, Modeling and Simulation**<br>**Session TM-TuM**<br>**Characterization & Modelling III**<br>**Moderator:**<br>**Michael Scarpulla**, University of Utah |
| 9:45am | **TM-TuM-6** Theory of Acceptor-Donor Complexes in $Ga_2O_3$, *I. Chatratin, F. Sabino,* University of Delaware; *P. Reunchan,* Kasetsart University, Thailand; ***Anderson Janotti***, University of Delaware | |
| 10:00am | **TM-TuM-7** Donor Doping of Monoclinic and Corundum $(Al_xGa_{1-x})_2O_3$, ***Darshana Wickramaratne***, US Naval Research Laboratory; *J. Varley,* Lawrence Livermore National Laboratory; *J. Lyons,* US Naval Research Laboratory | |
| 10:15am | **TM-TuM-8** The Co-Design, Fabrication, and Characterization of a Ga2O3-on-SiC MOSFET, ***Yiwen Song***, Pennsylvania State University; *A. Bhattacharyya,* University of Utah; *A. Karim, D. Shoemaker,* Pennsylvania State University; *H. Huang,* Ohio State University; *C. McGray,* Modern Microsystems, Inc.; *J. Leach,* Kyma Technologies, Inc.; *J. Hwang,* Ohio State University; *S. Krishnamoorthy,* University of California at Santa Barbara; *S. Choi,* Pennsylvania State University | |
| 10:30am | **BREAK** | |
| 10:45am | **INVITED: AC-TuM-10** Defects in Gallium Oxide – How We "See" and Understand Them, ***Jinwoo Hwang***, The Ohio State University | **Advanced Characterization Techniques**<br>**Session AC-TuM**<br>**Advanced Characterization & Microscopy**<br>**Moderator:**<br>**Ginger Wheeler**, Naval Research Laboratory |
| 11:15am | **AC-TuM-12** Atomic-Scale Investigation of Point and Extended Defects in Ion Implanted β-$Ga_2O_3$, ***Hsien-Lien Huang***, *C. Chae,* The Ohio State University; *A. Senckowski, M. Wong,* Penn State University; *J. Hwang,* The Ohio State University | |
| 11:30am | **AC-TuM-13** Microscopic and Spectroscopic Analysis of (100), (-201) and (010) $(Al_xGa_{1-x})_2O_3$ Films Using Atom Probe Tomography, *J. Sarker,* University at Buffalo-SUNY; *A. Bhuiyan, Z. Feng, L. Meng, H. Zhao,* The Ohio State University; ***Baishakhi Mazumder***, University at Buffalo-SUNY | |
| 11:45am | **AC-TuM-14** Phase and Microstructure Evolution of κ-$Ga_2O_3$ Thin Films Grown by MOCVD, ***Jingyu Tang***, *K. Jiang,* Carnegie Mellon University, China; *M. Cabral, A. Park,* Carnegie Mellon University; *L. Gu,* Carnegie Mellon University, China; *R. Davis , L. Porter,* Carnegie Mellon University | |
| 12:00pm | **AC-TuM-15** Investigation of Extended Defects in Ga2O3 Substrates and Epitaxial Layers using X-ray Topography, ***Nadeemullah A. Mahadik***, *M. Tadjer, T. Anderson, K. Hobart,* Naval Research Laboratory, USA; *K. Sasaki, A. Kuramata,* Novel Crystal Technology, Japan | |

# Tuesday Afternoon, August 9, 2022

| | Room Jefferson 2-3 | |
|---|---|---|
| 1:45pm | **INVITED: EG-TuA-1** Progress in Beta-Gallium Oxide Materials and Properties, ***James Speck***, University of California Santa Barbara | **Epitaxial Growth**<br>**Session EG-TuA**<br>**Bulk & Epitaxy II**<br>**Moderator:**<br>**Xiuling Li**, University of Texas Austin |
| 2:15pm | **EG-TuA-3** (110) β-$Ga_2O_3$ Epitaxial Films Grown by Plasma-Assisted Molecular Beam Epitaxy, ***Takeki Itoh***, *A. Mauze, Y. Zhang, J. Speck,* University of California at Santa Barbara | |
| 2:30pm | **EG-TuA-4** Si-doped β-Ga2O3 Films Grown at 1 µm/hr by Suboxide MBE, ***Kathy Azizie***, *P. Vogt, F. Hensling, D. Schlom, J. McCandless, H. Xing, D. Jena,* Cornell University; *D. Dryden, A. Neal, S. Mou, T. Asel, A. Islam, A. Green, K. Chabak,* Air Force Research Laboratory | |
| 2:45pm | **INVITED: EG-TuA-5** MOCVD Growth of $Ga_2O_3$ and $(Al_xGa_{1-x})_2O_3$, ***Hongping Zhao***, The Ohio State University | |
| 3:15pm | | |
| 3:30pm | **BREAK** | |
| 3:45pm | **DI-TuA-9** Dielectric Integration on (010) β-$Ga_2O_3$: $Al_2O_3$, $SiO_2$ Interfaces and their Thermal Stability, ***Ahmad Islam***, Air Force Research Laboratory; *A. Miesle,* University of Dayton; *M. Dietz,* Wright State University; *K. Leedy, S. Ganguli,* Air Force Research Laboratory; *G. Subramanyam,* University of Dayton; *W. Wang,* Wright State University; *N. Sepelak, D. Dryden,* KBR, Inc.; *T. Asel, A. Neal, S. Mou, S. Tetlak, K. Liddy, A. Green, K. Chabak,* Air Force Research Laboratory | **Dielectric Interfaces**<br>**Session DI-TuA**<br>**Processes & Devices II**<br>**Moderator:**<br>**Hongping Zhao,** Ohio State University |
| 4:00pm | **DI-TuA-10** Deep Etch Field-Terminated β-$Ga_2O_3$ Schottky Barrier Diodes With 4.2 MV/cm Parallel Plate Field Strength, ***Sushovan Dhara***, *N. Kalarickala, A. Dheenan, C. Joishi, S. Rajan,* The Ohio State University | |
| 4:15pm | **DI-TuA-11** Demonstration of Low Thermal Resistance in $Ga_2O_3$ Schottky Diodes by Junction-Side-Cooled Packaging, ***Boyan Wang***, *M. Xiao, J. Knoll, Y. Qin,* Virginia Polytechnic Institute and State University; *J. Spencer,* U.S. Naval Research Laboratory; *M. Tadjer,* U.S. Naval Research Laboratory; *C. Buttay,* Univ Lyon, CNRS, INSA Lyon, Université Claude Bernard Lyon 1, Ecole Centrale de Lyon, Ampère, France; *K. Sasaki,* Novel Crystal Technology, Japan; *G. Lu, C. DiMarino, Y. Zhang,* Virginia Polytechnic Institute and State University | |
| 4:30pm | **DI-TuA-12** High Temperature In-situ MOCVD-grown $Al_2O_3$ Dielectric on (010) β-$Ga_2O_3$ with 10 MV/cm Breakdown Field, ***Saurav Roy***, University of California Santa Barbara; *A. Bhattacharyya,* University of Utah; *C. Peterson, S. Krishnamoorthy,* University of California Santa Barbara | |
| 4:45pm | **DI-TuA-13** Metal Oxide (PtOX) Schottky Contact with High-k Dielectric Field Plate for Improved Field Management in Vertical β-Ga2O3 Devices, ***Esmat Farzana***, University of California Santa Barbara; *A. Bhattacharyya,* The University of Utah; *T. Itoh, S. Krishnamoorthy, J. Speck,* University of California Santa Barbara | |
| 5:00pm | **DI-TuA-14** Field Plated β-$Ga_2O_3$ Mis Diodes with High-κ $Tio_2$ Interlayer for Increased Breakdown and Reduced Leakage Current, ***Nolan Hendricks***, Air Force Research Laboratory; UC Santa Barbara; *A. Green, A. Islam, K. Leedy, K. liddy, J. Williams,* Air Force Research Lab; *E. Farzana, J. Speck,* UC Santa Barbara; *K. Chabak,* Air Force Research Lab | |

## Epitaxial Growth
## Room Jefferson 1 & Atrium - Session EG-TuP
## Epitaxial Growth Poster Session
## 5:15pm – 7:15pm

**EG-TuP-1** α-phase Gallium Oxide Thin Films Stabilized on a-, r- and m-plane Sapphire Substrates via Reactive Magnetron Sputtering and Pulsed Laser Deposition, ***Edgars Butanovs,*** Institute of Solid State Physics University of Latvia

**EG-TuP-2** Epitaxial Growth of $(Al_xGa_{1-x})_2O_3$ by Suboxide MBE, ***Jacob Steele****, K. Azizie, J. McCandless,* Cornell University*; T. Asel,* Air Force Research Lab*; H. Xing, D. Jena, D. Schlom,* Cornell University

**EG-TuP-3** LPCVD Grown n-$Ga_2O_3$ on p-GaN and Demonstration of p-n Heterojunction Behavior, ***Arnab Mondal****, A. Nandi, M. Yadav,* Indian Institute of Technology Mandi, India*; A. Bag,* Indian Institute of Technology Guwahati, India

**EG-TuP-5** Free Carrier Control in Homoepitaxial β-$Ga_2O_3$ Thin Films by Tin Impurity Doping, ***Neeraj Nepal****, B. Downey, V. Wheeler, D. Katzer, E. Jin, M. Hardy, V. Gokhale, T. Growden,* US Naval Research Laboratory*; K. Chabak,* Air Force Research Laboratory*; D. Meyer,* US Naval Research Laboratory

**EG-TuP-6** MBE Growth of Doped and Insulating Homoepitaxial β-$Ga_2O_3$, ***Jon McCandless****, V. Protasenko, B. Morell,* Cornell University*; E. Steinbrunner, A. Neal,* Air Force Research Laboratory, Materials and Manufacturing Directorate, USA*; Y. Cho, N. Tanen, H. Xing, D. Jena,* Cornell University

**EG-TuP-7** High Conductivity Homoepitaxial β-$Ga_2O_3$ Regrowth Layers by Pulsed Laser Deposition, ***Hyung Min Jeon****,* KBR*; K. Leedy,* Air Force Research Laboratory

**EG-TuP-8** Low-Temperature Epitaxial Growth and in Situ Atomic Layer Doping of β-$Ga_2O_3$ Films via Plasma-Enhanced ALD, ***Saidjafarzoda Ilhom****, A. Mohammad, J. Grasso, B. Willis,* University of Connecticut*; A. Okyay,* Stanford University*; N. Biyikli,* University of Connecticut

**EG-TuP-9** Highly conductive β-$Ga_2O_3$ and $(Al_xGa_{1-x})_2O_3$ epitaxial films by MOCVD, ***Fikadu Alema****,* Agnitron Technology*; T. Itoh, J. Speck,* Materials Department, University of California, Santa Barbara*; A. Osinsky,* Agnitron Technology

## Material and Device Processing and Fabrication Techniques
## Room Jefferson 1 & Atrium - Session MD-TuP
## Material and Device Processing and Fabrication Techniques Poster Session – 5:00pm – 7:00pm

**MD-TuP-1** Record Low Specific Resistance Ohmic Contacts to Highly Doped MOVPE-Grown β-$Ga_2O_3$ and β-$(Al_xGa_{1-x})_2O_3$ Epitaxial Films, ***Carl Peterson****,* University of California Santa Barbara*; F. Alema,* Agnitron Technology*; S. Roy,* University of California Santa Barbara*; A. Bhattacharyya,* University of Utah*; A. Osinsky,* Agnitron Technology*; S. Krishnamoorthy,* University of California Santa Barbara

**MD-TuP-3** MOCVD β-$Ga_2O_3$ Gate-recessed MESFET, ***Hannah Masten****, J. Lundh, J. Spencer,* US Naval Research Laboratory*; F. Alema, A. Osinsky,* Agnitron Technology*; A. Jacobs, K. Hobart, M. Tadjer,* US Naval Research Laboratory

**MD-TuP-4** Subsurface Damage Analysis of Chemical Mechanical Polished (010) β-$Ga_2O_3$ Substrates, ***Michael Liao****, K. Huynh, L. Matto, D. Luccioni, M. Goorsky,* UCLA

**MD-TuP-5** Diffusion of Zn in β-$Ga_2O_3$, ***Ylva Knausgård Hommedal****, Y. Frodason, L. Vines, K. Johansen,* Centre for Materials Science and Nanotechnology/Dep. of Physics, University of Oslo, Norway

**MD-TuP-6** Initial Nucleation of Metastable γ-$Ga_2O_3$ During sub-Millisecond Thermal Anneals of Amorphous $Ga_2O_3$ , ***Katie Gann****, C. Chang, M. Chang, D. Sutherland, A. Connolly, D. Muller, R. van Dover, M. Thompson,* Cornell University

**MD-TuP-7** Heavily Doped β-$Ga_2O_3$ Deposited by Magnetron Sputtering, ***Adetayo Adedeji****,* Elizabeth City State University*; J. Lawson, C. Ebbing,* University of Dayton Research Institute*; J. Merrett,* Air Force Research Laboratory

**MD-TuP-8** Point Defect Distributions in Ultrafast Laser Induced Periodic Surface Structures on β-$Ga_2O_3$, *D. Ramdin, E. DeAngelis, M. Noor, M. Haseman, E. Chowdhury,* ***Leonard Brillson****,* Ohio State University

## Theory, Modeling and Simulation
## Room Jefferson 1 & Atrium - Session TM-TuP
## Theory, Modeling and Simulation Poster Session
## 5:00pm - 7:00pm

**TM-TuP-1** Simulation Study of Single Event Effects in $Ga_2O_3$ Schottky Diodes, ***Animesh Datta****, U. Singisetti,* University at Buffalo

**TM-TuP-2** Anisotropic Photoresponsivity and Deviation from Beer-Lambert Law in Beta Gallium Oxide, ***Md Mohsinur Rahman Adnan****, D. Verma, S. Dhara,* The Ohio State University*; C. Sturm,* Universitat Leipzig, Germany*; S. Rajan, R. Myers,* The Ohio State University

**TM-TuP-4** Self-Trapped Holes and Polaronic Acceptors in Ultrawide Bandgap Oxides, ***John Lyons****,* US Naval Research Laboratory

**TM-TuP-5** Modeling for a High-Temperature Ultra-Wide Bandgap Gallium Oxide Power Module, ***Benjamin Albano****,* Virginia Tech Center for Power Electronics Systems*; B. Wang, C. DiMarino, Y. Zhang,* Virginia Tech Center for Power Electronics

**TM-TuP-6** Atomic Surface Structure of Sn doped β-$Ga_2O_3$(010) Studied by Low-energy Electron Diffraction, ***Alexandre Pancotti****,* Universidade Federal de Jataí, Brazil*; J. T. Sadowski,* Center for Functional Nanomaterials, Brookhaven National Laboratory*; A. Sandre Kilian,* Universidade Federal de Jataí, Brazil*; D. Duarte dos Reis,* Universidade Federal do Mato Grosso do Sul, Brazil*; C. Lubin,* SPEC, CEA, CNRS, Université Paris-Saclay, CEA Saclay , France*; A. Boucly,* SPEC, CEA, CNRS, Université Paris-Saclay, France*; P. Soukiassian,* SPEC, CEA, CNRS, Université Paris-Saclay, CEA Saclay, France*; J. Boeckl, D. Dorsey ,* Air Force Research Laboratory*; M. Shin, T. ASEL,* Air Force Research Lab*; J. Brown, N. Barrett ,* SPEC, CEA, CNRS, Université Paris-Saclay, CEA Saclay , France*; T. Back,* SPEC, CEA, CNRS, Université Paris-Saclay, CEA Saclay

# Wednesday Morning, August 10, 2022

| | Room Jefferson 2-3 | |
|---|---|---|
| 8:30am | **Welcome and Sponsor Thank Yous** | **Plenary Session**<br>**Session PS2-WeM**<br>**Plenary Session II**<br>**Moderator:**<br>**Kelson Chabak**, Air Force Research Laboratory |
| 8:45am | **INVITED: PS2-WeM-2** Plenary Lecture: Fundamental Limits of $Ga_2O_3$ Power Devices and How to Get There, ***Huili Grace Xing***, Cornell University | |
| 9:15am | **EP-WeM-4** Remarkable Improvement of Conductivity in Β-$Ga_2O_3$ by High-Temperature Si Ion Implantation, ***Arka Sardar**, T. Isaacs-Smith, S. Dhar,* Auburn University; *J. Lawson, N. Merrett,* Air Force Research Laboratory, USA | **Electronic and Photonic Devices, Circuits and Applications**<br>**Session EP-WeM**<br>**Process & Devices III**<br>**Moderator:**<br>**Uttam Singisetti**, University of Buffalo, SUNY |
| 9:30am | **INVITED: EP-WeM-5** Towards Lateral and Vertical $Ga_2O_3$ Transistors for High Voltage Power Switching, ***Kornelius Tetzner**, J. Würfl, E. Bahat-Treidel, O. Hilt,* Ferdinand-Braun-Institut, Leibniz-Institut für Höchstfrequenztechnik (FBH), Germany; *Z. Galazka, S. Bin Anooz, A. Popp,* Leibniz-Institut für Kristallzüchtung (IKZ), Germany | |
| 10:00am | **EP-WeM-7** Comparison of β-$Ga_2O_3$ Mosfets With TiW and NiAu Metal Gates for High-Temperature Operation, ***Nicholas Sepelak***, KBR, Wright State University; *D. Dryden,* KBR; *R. Kahler,* University of Texas at Dallas; *J. William,* Air Force Research Lab, Sensors Directorate; *T. Asel,* Air Force Research Laboratory, Materials and Manufacturing Directorate; *H. Lee,* University of Illinois at Urbana-Champaign; *K. Gann,* Cornell University; *A. Popp,* Leibniz-Institut für Kristallzüchtung, Germany; *K. Liddy,* Air Force Research Lab, Sensors Directorate; *K. Leedy,* Air Force Research Laboratory, Sensors Directorate; *W. Wang,* Wright State University; *W. Zhu,* University of Illinois at Urbana-Champaign; *M. Thompson,* Cornell University; *S. Mou,* Air Force Research Laboratory, Materials and Manufacturing Directorate, USA; *K. Chabak, A. Green,* Air Force Research Laboratory, Sensors Directorate; *A. Islam,* Air Force Research Laboratory, Sensors Directory | |
| 10:15am | **EP-WeM-8** High Electron Mobility Si-doped β-$Ga_2O_3$ MESFETs, ***Arkka Bhattacharyya***, University of Utah; *S. Roy,* University of California at Santa Barbara; *P. Ranga,* University of Utah; *S. Krishnamoorthy,* University of California at Santa Barbara | |
| 10:30am | **BREAK** | |
| 10:45am | **EP2-WeM-10** *β*-$Ga_2O_3$ Lateral FinFETs Formed by Atomic Ga Flux Etching, ***Ashok Dheenan**, N. Kalarickal, Z. Feng, L. Meng,* The Ohio State University; *A. Fiedler,* IKZ Berlin, Germany; *C. Joishi, A. Price, J. McGlone, S. Dhara, S. Ringel, H. Zhao, S. Rajan,* The Ohio State University | **Electronic and Photonic Devices, Circuits and Applications**<br>**Session EP2-WeM**<br>**Process and Devices IV**<br>**Moderator:**<br>**Christina DiMarino**, Virginia Tech |
| 11:00am | **EP2-WeM-11** Insights Into the Behaviour of Leakage Current in Lateral $Ga_2O_3$ Transistors on Semi-Insulating Substrates, ***Zequan Chen**, A. Mishra, M. Smith, T. Moule,* University of Bristol, UK; *M. Uren,* University of Brsitol, UK; *S. Kumar, M. Higashiwaki,* National Institute of Information and Communications Technology, Japan; *M. Kuball,* University of Bristol, UK | |
| 11:15am | **EP2-WeM-12** Device Figure of Merit Performance of Scaled Gamma-Gate β-$Ga_2O_3$ MOSFETs, ***Kyle Liddy**, A. Islam, J. Williams, D. Walker, N. Moser, D. Dryden, N. Sepelak, K. Chabak, A. Green,* AFRL | |
| 11:30am | **EP2-WeM-13** Electromigration of Native Point Defects and Breakdown in $Ga_2O_3$ Vertical Devices, *M. Haseman, D. Ramdin,* Ohio State University; *W. Li, K. Nomoto, D. Jena, G. Xing,* Cornell University; ***Leonard Brillson**,* Ohio State University | |
| 11:45am | **Closing Remarks, Sponsor Thank Yous, & Collection of e-Surveys** | |

# Author Index

## Bold page numbers indicate presenter

# Author Index

## — S —

## — T —

## — U —

## — V —

## — W —

## — X —

## — Y —

## — Z —